\documentclass[usegraphicx,usenatbib,useAMS]{mn2e}
\usepackage{times}
\newcommand{\be}{\begin{equation}}
\newcommand{\ba}{\begin{eqnarray}}
\newcommand{\ee}{\end{equation}}
\newcommand{\ea}{\end{eqnarray}}  

\def\lesssim{\mathrel{\hbox{\rlap{\hbox{\lower4pt\hbox{$\sim$}}}\hbox{$<$}}}}
\def\gtrsim{\mathrel{\hbox{\rlap{\hbox{\lower4pt\hbox{$\sim$}}}\hbox{$>$}}}}

\def\gtsima{$\; \buildrel > \over \sim \;$}
\def\ltsima{$\; \buildrel < \over \sim \;$}
\def\gsim{\lower.5ex\hbox{\gtsima}}
\def\lsim{\lower.5ex\hbox{\ltsima}}
\def\simgt{\lower.5ex\hbox{\gtsima}}
\def\simlt{\lower.5ex\hbox{\ltsima}}
\def\simpr{\lower.5ex\hbox{\prosima}}

\def\simless{\mathbin{\lower 3pt\hbox
   {$\rlap{\raise 5pt\hbox{$\char'074$}}\mathchar''7218$}}}   % < or of order
\def\simgreat{\mathbin{\lower 3pt\hbox
   {$\rlap{\raise 5pt\hbox{$\char'076$}}\mathchar''7218$}}}   % > or of order

\begin{document}

\title[Simulating Cosmic Reionization at Large Scales]
{Simulating Cosmic Reionization at Large Scales I: the Geometry of
  Reionization} 
\author[I. T. Iliev, et al.]{I.~T.~Iliev
$^1$\thanks{e-mail: iliev@cita.utoronto.ca},
G.~Mellema$^{2,3}$, 
U.-L.~Pen$^1$, 
H.~Merz$^1$, 
P.~R.~Shapiro$^4$, 
M.~A.~Alvarez$^4$
\\
$^1$ Canadian Institute for Theoretical Astrophysics, University
  of Toronto, 60 St. George Street, Toronto, ON M5S 3H8, Canada\\
$^2$ ASTRON, P.O. Box 1, NL-7990 AA Dwingeloo, The
  Netherlands \\
$^3$ Sterrewacht Leiden, P.O. Box 9513, NL-2300 RA Leiden, The
  Netherlands \\
$^4$ Department of Astronomy, University of Texas,
  Austin, TX 78712-1083
}
\date{Revised 2005 December 4}
\pubyear{2005} \volume{000} \pagerange{1}
%\twocolumn
\maketitle\label{firstpage}
%\title{Simulating Cosmic Reionization at Large Scales I: the Geometry of
%  Reionization} \author{Ilian~T.~Iliev$^1$, Garrelt Mellema$^{2,3}$, Ue-Li
%  Pen$^1$, Hugh Merz$^1$, Paul.R. Shapiro$^4$, Marcelo A. Alvarez$^4$}

%\altaffiltext{1}{Canadian Institute for Theoretical Astrophysics, University
%  of Toronto, 60 St. George Street, Toronto, ON M5S 3H8, Canada}
%\altaffiltext{2}{Sterrewacht Leiden, P.O. Box 9513, NL-2300 RA Leiden, The
%  Netherlands} 
%\altaffiltext{3}{ASTRON, P.O. Box 1, NL-7990 AA Dwingeloo, The
%  Netherlands} 
%\altaffiltext{4}{Department of Astronomy, University of Texas,
%  Austin, TX 78712-1083}
%\label{firstpage}

\begin{abstract}
  We present the first large-scale radiative transfer simulations of cosmic 
  reionization, in a simulation volume of $(100\,h^{-1}\rm Mpc)^3$. This
  is more than a 2 orders of magnitude improvement over previous simulations. 
  We achieve this by combining the results from extremely large, cosmological, 
  N-body simulations with a new, fast and efficient code for 3D radiative 
  transfer, $\rm C^2$-Ray, which we have recently developed. These simulations
  allow us to do the first numerical studies of the large-scale structure 
  of reionization which at the same time, and crucially, properly take 
  account of the dwarf galaxy ionizing sources which are primarily 
  responsible for reionization.
  In our realization, reionization starts around $z\sim21$, and final overlap
  occurs by $z\sim11$. The resulting electron-scattering optical depth is
  in good agreement with the first-year WMAP polarization data. We show 
  that reionization clearly proceeded in an inside-out fashion, with the 
  high-density regions being ionized earlier, on average, than the voids. 
  Ionization histories of smaller-size (5 to 10 comoving Mpc) subregions  
  exabit a large scatter about the mean and do not describe the global
  reionization history well. This is true even when these subregions are 
  at the mean density of the universe, which shows that small-box
  simulations of reionization have little predictive power for the evolution
  of the mean ionized fraction. The minimum reliable volume size for such 
  predictions is $\sim30$ Mpc. We derive the power-spectra of the neutral, 
  ionized and total gas density fields and show that there is a significant 
  boost of the density fluctuations in both the neutral and the ionized 
  components relative to the total at
  arcminute and larger scales. We find two populations of H~II regions according 
  to their size, numerous, mid-sized ($\sim10$ Mpc) regions and a few, rare, very 
  large regions tens of Mpc in size. Thus, local overlap on fairly large scales 
  of tens of Mpc 
  is reached by $z\sim13$, when our volume is only about 50\% ionized, and well 
  before the global overlap. We derive the statistical distributions of the 
  ionized fraction and ionized gas density at various scales and for the first 
  time show that both distributions are clearly non-Gaussian. 
  All these quantities are critical for predicting 
  and interpreting the observational signals from reionization from a variety 
  of observations like 21-cm emission, Ly-$\alpha$ emitter statistics, Gunn-Peterson
  optical depth and small-scale CMB secondary anisotropies due to patchy reionization.   
\end{abstract}

\section{Introduction}
Understanding the large-scale geometry of reionization (sometimes also
referred to as topology of reionization), i.e.\ the size- and 
spatial distribution of the ionized and neutral patches, and
their evolution in time is one of the most important problems in this
fast-developing field.  Better understanding of this
geometry is crucial for making detailed and realistic predictions for
the observational features of reionization. Detecting these features is the
goal of a number of upcoming meter-wavelength radio synthesis
telescopes, such as PAST\footnote{\tt
http://web.phys.cmu.edu/$\!\sim\!$past/}, LOFAR\footnote{\tt
http://www.lofar.org}, MWA\footnote{\tt
http://web.haystack.mit.edu/arrays/MWA}, and SKA\footnote{\tt
http://www.skatelescope.org}. Also being planned are observations of
small-scale CMB anisotropies created by ionized patches
\citep[e.g.][]{2003ApJ...598..756S}, and direct observations of
high-redshift Ly-$\alpha$ emitters and their clustering using either
James Webb Space Telescope or ground-based telescopes 
\citep[e.g.][]{2004ApJ...617L...5M,2005ApJ...619...12S}. Such
observations can in principle map the complete progress of
reionization through time and space.

%The reionization of the universe, from the formation of the first stars at
%high redshift ($z>30$) until the almost complete ionization of the
%intergalactic medium (IGM) by $z\sim6$ demonstrated by the lack of
%Gunn-Peterson trough in QSO and galactic spectra, is one of the most
%fascinating but least well-understood epochs in the history of the universe. 
%The first ground-breaking observations indicated that the reionization 
%process
%was well-advanced quite early ($z>15$), but that the final overlap occurred
%much later, around $z\sim6$.   

In the last few years a variety of different cosmological radiative transfer
methods have been developed. These generally fall into two basic groups,
moment methods
\citep{2001NewA....6..437G,2002ApJS..141..211C,2003ApJS..147..197H}, and
ray-tracing methods \citep{1998A&A...331..335M,1999MNRAS.309..287R,
  1999ApJ...523...66A,2000MNRAS.314..611C,2001MNRAS.321..593N,
  2001NewA....6..359S,2002ApJ...572..695R,2003A&A...405..189L,
  2003MNRAS.345..379M,2004MNRAS.348..753S,
  2004MNRAS.348L..43B,2005MNRAS...361..405I,methodpaper}.
Several simulations of reionization have been performed using some of these 
codes, most often as a post-processing step  (i.e. the ``static limit'' which
neglects the gasdynamical response to photoionization and heating)  
to cosmological N-body simulations with and without gas 
\citep{2001MNRAS.321..593N,2001NewA....6..359S,2002ApJ...572..695R,
2003MNRAS.345..379M},
while others directly coupled the radiative transfer to the gas evolution and
followed the evolution self-consistently \citep{2001NewA....6..437G}. Despite
these significant advances, all current reionization simulations are limited
to fairly small volumes, with computational box sizes not exceeding
$20\,h^{-1}$ comoving Mpc, and usually much smaller than that. 
% Such small volumes do not allow for studies of the large-scale geometry 
%of reionization and for realistic observational predictions. 
%Detailed knowledge of the early nonlinear structures forming in a $\Lambda$CDM
%universe is the first key component in reionization studies.  
The main reason for this limitation is that the ionizing
photon output during reionization is dominated by dwarf galaxies, which at
early times are far more numerous than the larger galaxies, while the ionizing
photon consumption (ionizations and recombinations) is dominated by even
smaller structures, due to the hierarchical nature of $\Lambda$CDM structure
formation. The need to resolve such small galaxies imposes a severe limit on
the computational box size. On the other hand, these ionizing sources were
strongly clustered at high redshift and, as a consequence, the H~II regions
they created are expected to quickly overlap and grow to very large 
sizes, reaching up to tens of Mpc
\citep[e.g.][]{2004ApJ...609..474B,2005MNRAS.363.1031F,2005astro.ph..7014C,2005pgqa.conf..369I}.  
The many orders of magnitude of difference between these scales
demand extremely high resolution from any simulations designed to study early
structure formation from the point of view of reionization. Further limitations
are imposed by the low efficiency of the used radiative transfer methods. Most
methods are based on ray-tracing and thus are quite accurate, but their
computational expense grows roughly proportionally to the number of ionizing
sources present. This generally renders them impractical when more than a few
thousand ionizing sources are involved, severely limiting the computational
volume that can be simulated effectively.   

%Studies of the geometry of reionization - the distribution of the ionized and
%neutral regions in sizes and shapes both in space and time - have long been
%hampered by too small simulation box sizes. The forming H~II regions are quite
%large, of order up to tens of comoving Mpc, requiring computational volumes of
%at least that size for correct modelling and fair statistical sampling.
%The details of the reionization geometry are important for predicting correctly 
%the various observational signatures of the reionization epoch, e.g. redshifted 
%21-cm emission, secondary CMB anisotropies at small scales, and high-redshift
%Ly-$\alpha$ emitter statistics.

Over the years many analytical approaches to modelling reionization have been 
proposed
\citep[e.g.][]{1994ApJ...427...25S,2003ApJ...595....1H,2003ApJ...586..693W,2004ApJ...613....1F,
2005ApJ...624..491I}. However, these models are all statistically-averaged 
ones and can thus only make statistical predictions. Moreover, 
they have not been checked against simulations or observations and hence their 
level of reliability is currently unclear. In general, semi-analytical models 
are inevitably simplified in order to render the problem tractable and their 
prediction power depends on how well they can represent the relevant features
of reionization. They could be quite useful in situations when simulations are
too expensive, e.g. for exploring the parameter space, or studying very rare
objects which requires huge volumes, well beyond the reach of current
simulations. However, the correctness, reliability, and limitations of any
semi-analytical model should still be established first by comparison with
simulations.   

\begin{figure*}
\includegraphics[width=6.5in]{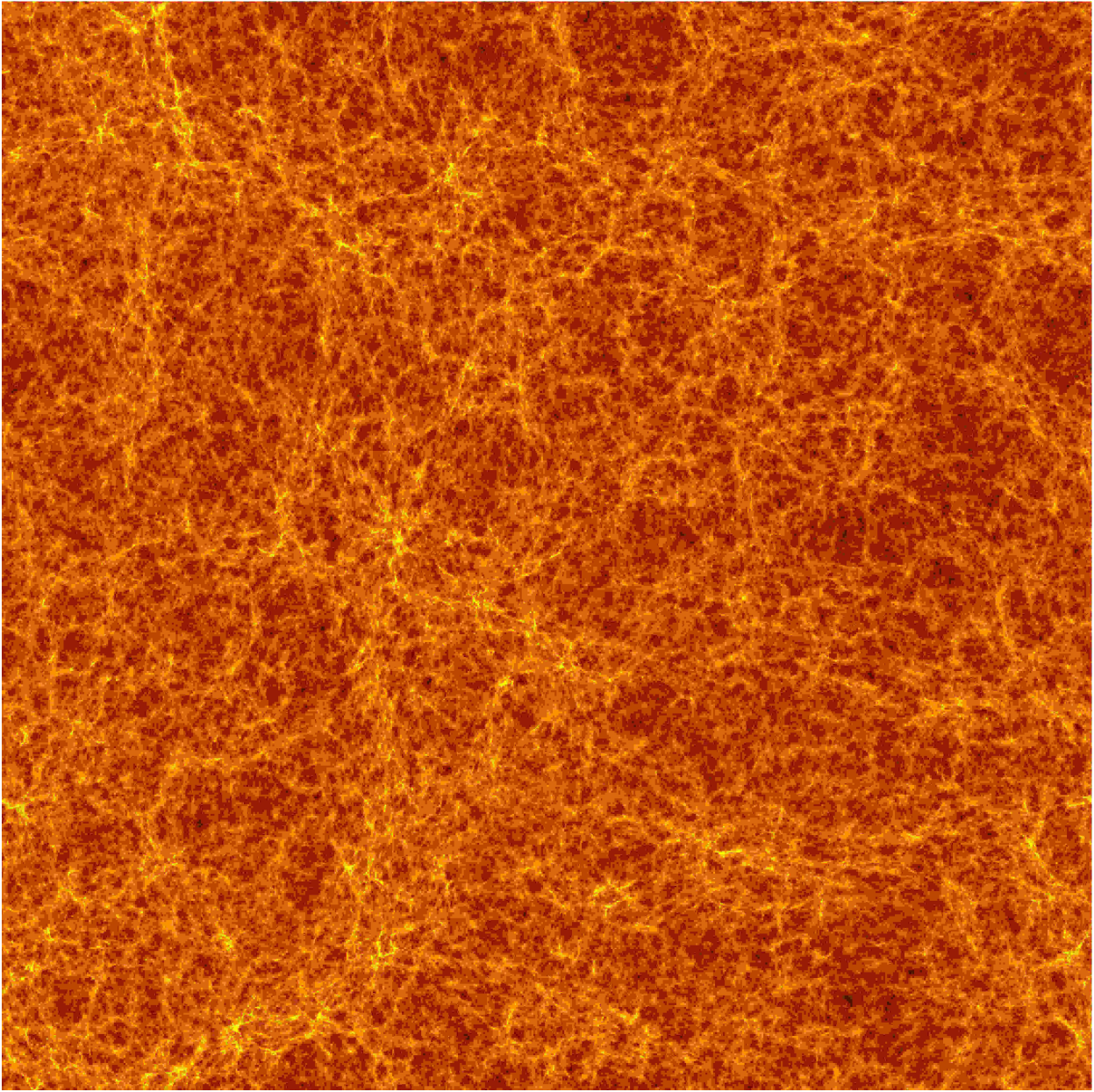}
\caption{Early structure formation in $\Lambda$CDM, at $z=10$, from our N-body 
  simulation: projection of the cloud-in-cell densities on the fine simulation grid
  ($3248\times3248$ pixels) in a 20 comoving Mpc slice ($\sim 6\times10^8$ particles 
  in the slice) of the $(100\,h^{-1})^3$ Mpc$^3$ simulation volume. (See $\tt
  http://www.cita.utoronto.ca/\!\sim\!iliev/research.html$ for the
  full-resolution images and some movies of our simulations).
\label{simul_fig}}
\end{figure*}

The current development of novel, more efficient codes for both
cosmological N-body and hydrodynamical simulations and for numerical radiative
transfer finally allows reionization simulations in large volumes. In this
work we present the first large-scale, in a volume of $(100\,h^{-1}\rm Mpc)^3$, 
radiative transfer simulations of this process. We achieve this by 
combining the results from an extremely large N-body simulation performed
with the code PMFAST \citep{2005NewA...10..393M} with a very fast and
efficient cosmological radiative transfer code called $C^2$-Ray which we 
have recently developed \citep{methodpaper}. Our underlying N-body simulation has a
sufficient mass resolution to resolve all halos down to dwarf
galaxies inside our volume reliably, as well as their clustering and the relevant
large-scale fluctuations of the density field. Our new  ray-tracing radiative
transfer method is based on explicit photon conservation in space and time
which allows us to use large time steps and fairly coarse grids without
loss of accuracy. Ionization fronts (I-fronts) are correctly tracked even for 
very optically-thick
cells. These features make our code far more efficient than previous ones,
allowing for faithful treatment (using parallel machines) of tens, even 
hundreds of thousands of ionizing sources on much larger grids than before.

We also note the very recent results of \citet{2005ApJ...633..552K,
2005astro.ph.11627K} which simulate extremely  large cosmological 
volumes, up to $\sim 1$ Gpc in size. However, these simulations have very 
coarse resolutions and do not resolve the individual ionizing sources.
These are instead represented only in a mean, approximate way based on
separate, much smaller scale radiative transfer simulations. Such approach 
may be appropriate for certain problems, like the one discussed in these 
papers, namely modelling the rare, bright high-redshift QSO's. However it 
cannot be used to answer the questions posed and addressed in the present
work, due to its approximate nature and lack of both a source population
resolution and a spatial one.

We assume that the gas is closely following the dark
matter distribution, which is a quite accurate assumption at the large scales
considered here \citep{2004MNRAS.355..451Z}. Furthermore, the gas back
reaction due to the ionization can be ignored to a good approximation at these
scales since the global, large-scale I-fronts are highly
supersonic, of weak R-type (from ``rarefied'', i.e. typically occurring in 
low-density gas), out-running any reaction of the gas
\citep{1987ApJ...321L.107S}. This approximation breaks down in dense gas
inside halos, where the I-fronts slow down and transform to a D-type
(from ``dense'', i.e. typically occurring in dense gas),
generally preceded by a shock 
\citep{2004MNRAS.348..753S,2005MNRAS...361..405I}. 
Thus, on smaller scales a fully-coupled hydrodynamic and radiative transfer
treatment is required. 

The general outlay of this paper is as follows. We describe our
numerical methodology in \S~\ref{sim_sect}. In \S~\ref{results_sect} we
present our results: on the 
globally-averaged quantities (e.g. ionized fraction, mean number of
recombinations per atom, electron scattering optical depth) in
\S~\ref{global_quantities_sect} and on the reionization geometry in
\S~\ref{geometry_sect}. Finally, we discuss our results in
\S~\ref{discuss_sect}. This paper concentrates on the geometry of
reionization, the corresponding implications for the observation of the 21-cm
signal will be presented in a companion paper \citep{21cmreionpaper}.

Throughout this study we assume a flat $\Lambda$CDM cosmology with parameters
($\Omega_m,\Omega_\Lambda,\Omega_b,h,\sigma_8,n)=(0.27,0.73,0.044,0.7,0.9,1)$
\citep{2003ApJS..148..175S}, where $\Omega_m$, $\Omega_\Lambda$, and
$\Omega_b$ are the total matter, vacuum, and baryonic densities in units of
the critical density, $h$ is the Hubble constant in units of 100
$\rm km\,s^{-1}Mpc^{-1}$, $\sigma_8$ is the standard deviation of linear density
fluctuations at present on the scale of $8 \rm h^{-1}{\rm Mpc}$, and $n$ is the
index of the primordial power spectrum. We use the CMBfast transfer function
\citep{1996ApJ...469..437S}.

\begin{figure*}
\begin{center}
  \includegraphics[width=4in]{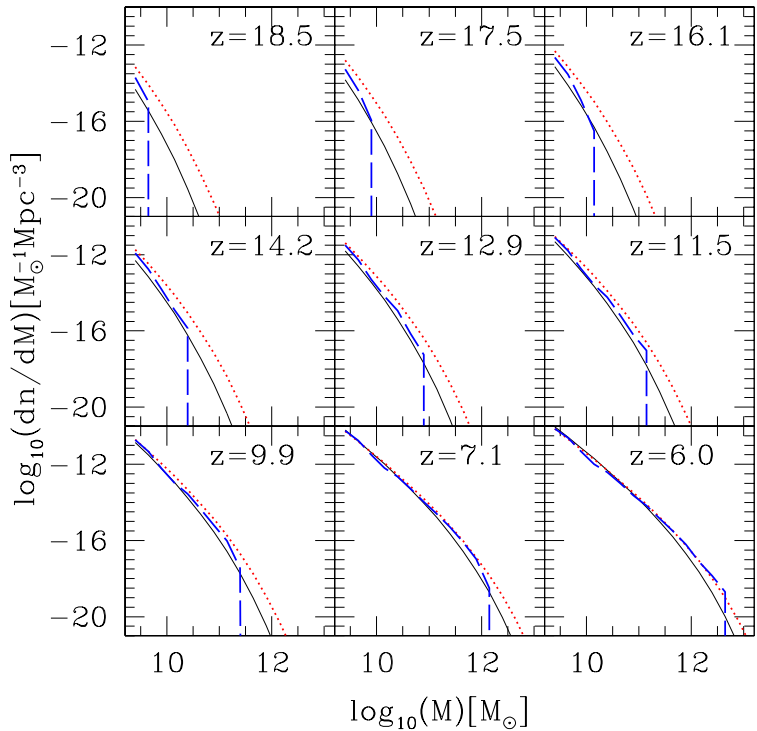}
\caption{Halo mass function from our simulation at various redshifts, as 
  labeled, (thick, long-dashed) and analytical approximations: the
  standard Press-Schechter (PS; thin, solid) and Sheth-Tormen (ST;
  thick, dotted).
\label{PS_fig}}
\end{center}
\end{figure*}

\section{The Simulations}
\label{sim_sect}

\subsection{N-body simulations}
We performed the underlying cosmological N-body simulations with the
particle-mesh cosmological code PMFAST \citep{2005NewA...10..393M} with
simulation volume of $(100\,\rm h^{-1} Mpc)^3$. Our resolution is
$1624^3=4.28$ billion dark matter 
particles and $3248^3$ computational cells (Figure~\ref{simul_fig}). The
particle mass is $2.5\times10^7M_\odot$ and in order to be conservative we 
consider only well-resolved halos which contain at least 100 particles. This 
gives a minimum halo mass of $2.5\times10^9M_\odot$, corresponding to dwarf 
galaxies. We find the halos and their detailed parameters ``on-the-fly'',
while the simulation is running, using a spherical overdensity method with 
overdensity of $\delta=130$ (although the results do not depend significantly 
on the particular overdensity value we have chosen). The first halos form at 
$z\approx21$ and the number of halos quickly rises thereafter, reaching over 
85,000 halos by $z\sim11$ and over 0.8 million halos by $z=6$. In 
Figure~\ref{PS_fig} we plot several sample halo mass functions from our
simulation at different redshifts. At lower redshifts ($z\lesssim10$) our 
mass function is in excellent agreement with the analytical Sheth-Tormen (ST)  
mass function \citep[e.g.][]{2002MNRAS.329...61S}, while at higher redshifts 
the ST mass function somewhat overestimates the actual number of halos. The 
standard Press-Schechter approximation \citep[PS;][]{1974ApJ...187..425P}, on
the other hand, significantly underestimates the number of rare halos in the 
exponential tail of the mass function but agrees fairly well with the
numerical mass function (as well as with the ST mass function) for less rare 
halos. Previously, \citet{2003MNRAS.346..565R} found a similar trend of ST 
overestimating the mass function at $z>10$, but being fairly accurate at later 
times.
%, and PS underestimating the number of rare halos at all times. 
We will present the detailed results from a series of these simulations with
several different computational volume sizes in a separate paper.

\begin{figure*}
\begin{center}
  \includegraphics[width=3in]{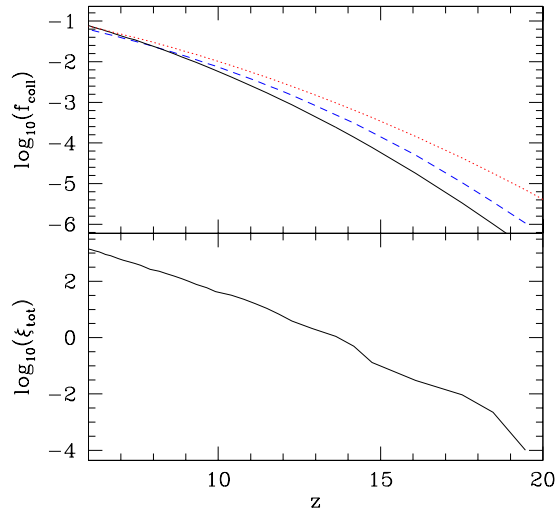}
\caption{(top) Collapsed mass fraction in halos in the simulation, $f_{\rm coll}$ 
  (dashed), and the analytical PS (solid) and ST (dotted) halo collapsed fractions 
  with the same minimum halo mass, and (bottom) cumulative number of ionizing photons 
  per hydrogen atom, $\xi_{\rm tot}$ emitted by all sources in the computational 
  volume.
\label{fcoll_fig}}
\end{center}
\end{figure*}

\subsection{Reionization simulations}
We performed our radiative transfer simulations using the cosmological
ray-tracing radiative transfer code $C^2$-Ray \citep{methodpaper}. The PMFAST
N-body simulation described above provided the density fields and the halo
catalogues containing the masses, positions and other detailed properties of the
collapsed halos. We have saved 50 of these density fields (of the $812^3$
coarse PMFAST density field), spaced at roughly equal time intervals, $\Delta
t_i\approx 20$ Myr, and the corresponding halo catalogues. The density field
is further re-gridded, to $406^3$ and $203^3$ computational cells, 
for radiative transfer runs at different resolutions.  We assume that the gas 
density distribution follows that of the dark matter. This is a quite
accurate assumption at the large scales considered in these simulations
\citep{2004MNRAS.355..451Z}.  In the interest of speed and manageability of the
calculations, we assume a fixed temperature of $10^4$ K everywhere. The effect 
of a varying temperature at these large scales would be to slightly modify the 
local recombination rate and will be studied in future work. Currently we employ
transmissive boundary conditions for the radiative transfer. While this leads 
to some loss of ionizing
photons through the boundaries, we have monitored the photon loss, and due to
our large simulation volume the effect never became significant until very
close to overlap. Subsequent versions of our code will include periodic
boundary conditions and more computationally-efficient handling of the
evolution of optically-thin gas, so we will be able to follow the pre- and
post-overlap evolution more precisely. We should also note that the observed 
IGM mean free path for hydrogen ionizing photons is much smaller than $100 h^{-1}$ 
Mpc comoving for $z>6$ \citep{2002AJ....123.1247F}, even when the IGM is
mostly ionized. This agrees with our results that only a small fraction of the
total ionizing photons emitted escapes the computational volume for the redshift 
range of interest in this paper, i.e. $z>11$.    

The ionizing sources in our simulations are based on the halos found in the
simulations. Sources falling within the same cell of the radiative transfer
grid are combined together and placed in the center of the cell.  
For simplicity we assume a constant mass-to-light ratio to assign 
ionizing flux to each halo, according to 
\be 
\dot{N}_\gamma=f_\gamma\frac{M\Omega_b}{\Delta t_i\Omega_0 m_p},
\label{ion_flux}
\ee 
where $\dot{N}_\gamma$ is the number of ionizing photons emitted by the
source per unit time, $M$ is the halo mass, and $m_p$ is the proton mass. The
efficiency factor $f_\gamma$ is defined by \be f_\gamma=f_*f_{\rm esc}N_i, \ee
where $N_i$ is the total photon production per stellar baryon, $f_*$ is the
star-formation efficiency and $f_{\rm esc}$ ionizing photon escape fraction.
Here we adopt the value $f_\gamma=2000$, equivalent to e.g.  $N_i=50,000$,
$f_*=0.2$ and $f_{\rm esc}=0.2$ (corresponding to a top-heavy IMF and moderate
star formation efficiency and escape fraction), or to $N_i=10,000$, $f_*=0.4$ and
$f_{\rm esc}=0.5$ (corresponding to a Salpeter IMF with high star formation
efficiency and escape fraction) \citep[see discussion and further references
in][]{2005ApJ...624..491I}. More sophisticated star formation models can be
adopted in the future. In Figure~(\ref{fcoll_fig}) we show the evolution
of the collapsed mass fraction in halos and the resulting cumulative number of
ionizing photons per atom $\xi$ emitted. The minimum required for completing 
reionization is one photon/atom and is reached at $z\sim13.5$. In
practice, however, each atom experiences recombinations during the
course of reionization and thus more than one photon per atom is
needed in order to complete the process. 

\begin{figure*}
\begin{center}
  \includegraphics[width=3in]{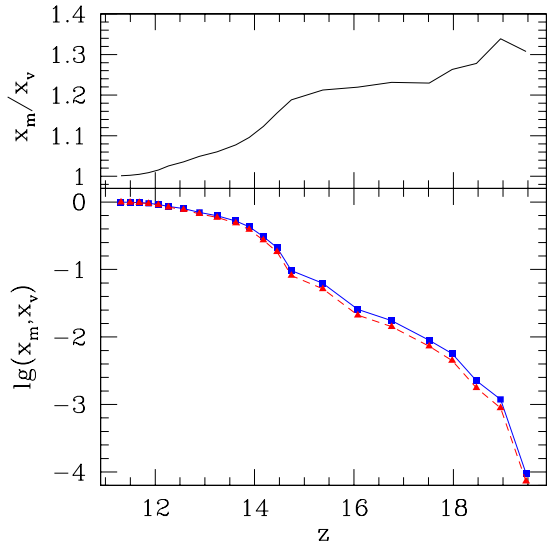}
\caption{(bottom) Evolution of the ionized fractions: mass-weighted, $x_m$, 
  (blue, solid, squares) and volume-weighted, $x_v$, (red, dashed, triangles); 
  and (top) their ratio, vs.  redshift $z$.
\label{fracs_fig}}
\end{center}
\end{figure*}
Our code is parallelized for SMP machines using OpenMP and runs highly
efficiently, fully utilizing all processors assuming there is sufficient
memory for all threads. This particular simulation took 2 weeks of computing
time on a Compaq machine with 32 alpha processors running at 733 MHz. At
$203^3$ resolution the code requires $\sim 0.4$ GB of RAM per computing thread
(12 GB total for 32 threads). The code also readily runs on a single- or
dual-processor workstation. On a single processor PC this simulation requires 
680 MB of RAM and would take about 2 months of computing time on
a current 64-bit Opteron workstation.   
    
\section{Results}
\label{results_sect}

\subsection{Globally-Averaged Quantities} 
\label{global_quantities_sect}

We start our analysis of the results by deriving from our simulation 
a number of globally-averaged characteristics of the reionization process. 
While such averaged quantities carry no direct information about 
the geometry of reionization, they do have important observational and 
theoretical implications.

\begin{figure*}
\begin{center}
  \includegraphics[width=3in]{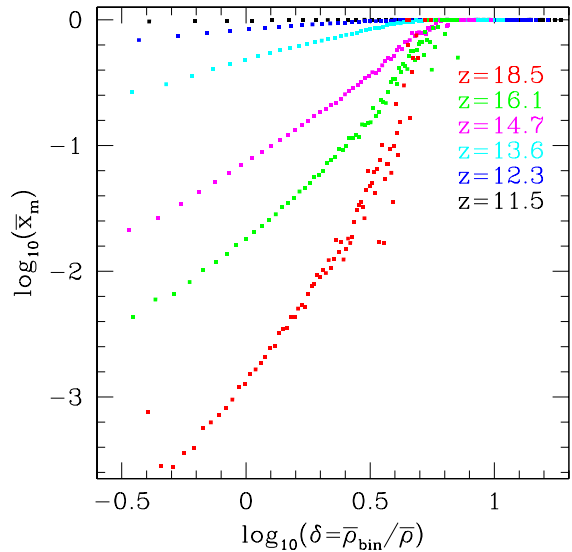}
  \caption{
A histogram of  $\bar{\rho}_{\rm bin}/\bar{\rho}$, the average ionized mass
fraction of the computational cells with a given overdensity, for
  several redshifts, as indicated (bottom to top from the highest to the
  lowest redshift, labeled also by color).\label{ion_dens_fig}}
\end{center}
\end{figure*}

\subsubsection{Mass- and volume weighted ionization fractions}
In our simulations the first ionizing sources form around $z\sim21$ and final
overlap (defined by less than 1\% neutral fraction remaining) is reached by 
$z\sim11.3$. The first
sources are highly clustered, in accordance with the Gaussian statistics of
high density peaks within which these first halos form, and are surrounded by
regions with density well above the cosmic mean. In Figure~(\ref{fracs_fig}) 
we show the evolution of mass-weighted ($x_m$) and volume-weighted ($x_v$) 
ionized fractions during the course of our simulation. The mass-weighted 
ionized fraction starts significantly higher, by 30-35\%, than the 
volume-weighted one. The difference between the two steadily decreases 
thereafter but remains above one throughout the evolution, eventually 
asymptoting to one when the whole computational volume becomes ionized. 
The ratio of the two ionized fractions, $x_m/x_v$, is in fact equal to the 
average gas density in the ionized regions in units of the mean density of 
the universe:
\be
\frac{x_m}{x_v}=
\frac{V_{\rm box}}{M_{\rm box}}\frac{x_mM_{\rm box}}{x_vV_{\rm box}}=
\frac1{\bar{\rho}}\frac{M_{\rm ionized}}{V_{\rm ionized}}
\ee
This is a manifestation of the predominantly inside-out 
character of reionization. The high-density regions and local density peaks 
surrounding the sources are ionized first. The ionization fronts 
then expand further into both high- and low-density nearby regions, with the
material in the large voids ionized last. This point is further illustrated in
Figure~\ref{ion_dens_fig}, 
where we show the histogram of the mean mass ionized fraction of all computational
cells in a given density bin (in units of the mean density) at several
redshifts, as indicated, covering the full range of interest. The
highest-density cells are almost instantly ionized and remain ionized
throughout the simulation, while the lower-density cells take progressively
longer to become ionized. Higher-density cells are on average always more
ionized than lower-density ones. Naturally, close to overlap the average
density of the ionized regions approaches the global mean density, and both
high- and low-density cells become mostly ionized. 
\begin{figure*}
\begin{center}
  \includegraphics[width=3in]{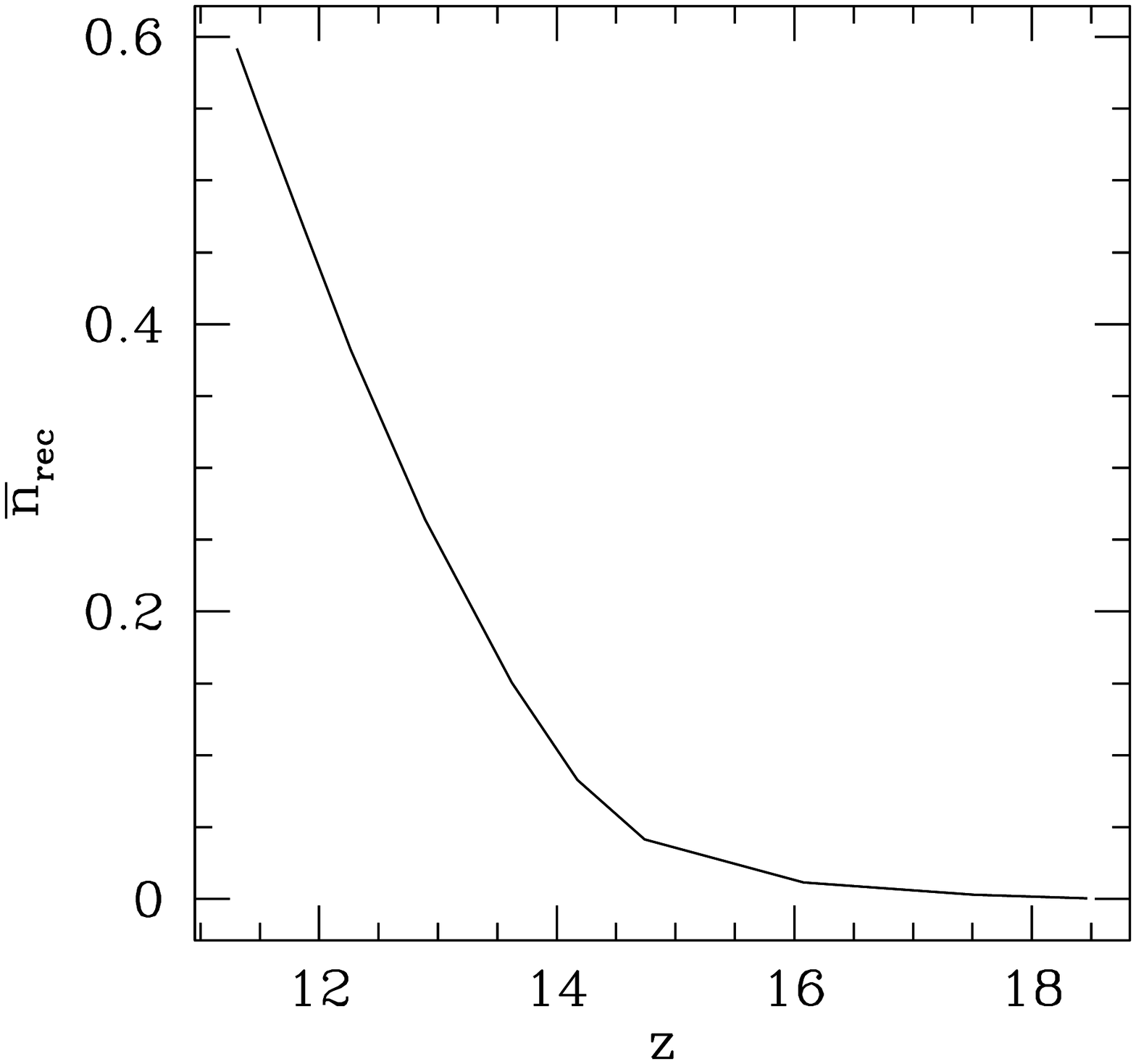}
  \includegraphics[width=3in]{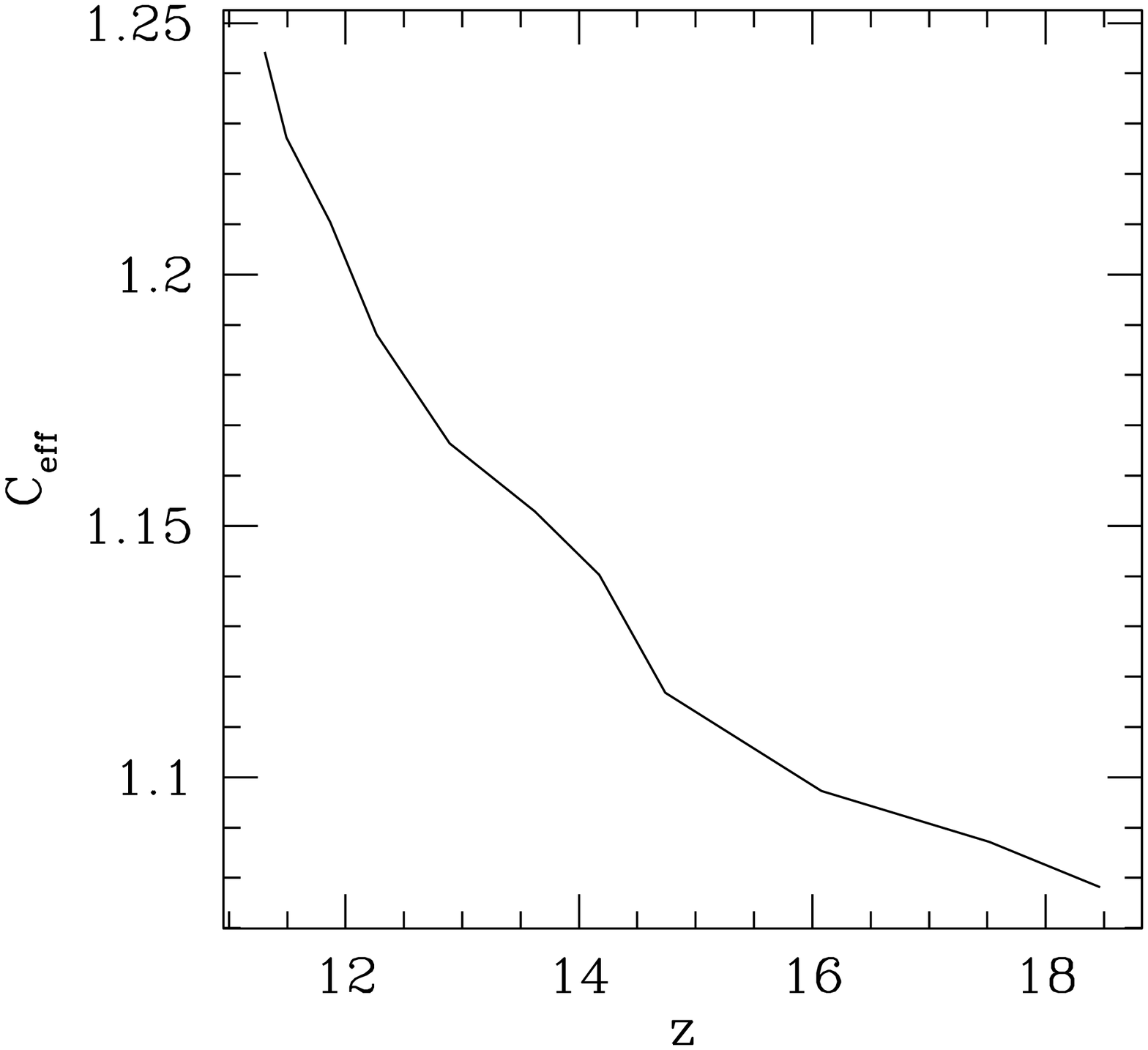}
\caption{(left) Time-integrated number of recombinations per total atom in our 
simulation volume, $\bar{n}_{\rm rec}$, vs. redshift. (right) Evolution of the 
effective clumping factor, 
$C_{\rm eff}=\langle n_{HII}^2\rangle/\langle n_{HII}\rangle^2$, of the ionized gas.
\label{ceff_fig}}
\end{center}
\end{figure*}

\subsubsection{Average number of recombinations and effective gas clumping factor}
In Figure~(\ref{ceff_fig}, left) we plot the time-integrated number of recombinations
per total atom in our simulation volume, $\bar{n}_{\rm rec}$, vs. redshift. This 
number of recombinations
starts low, since even within the denser regions which surround each source
the recombination time is relatively long. Around $z\sim14$ $\bar{n}_{\rm rec}$ 
quickly rises, eventually reaching $\sim 0.6$ at overlap -- i.e. each atom in 
the computational volume experienced on average 0.6 recombinations. This rise 
is due to the longer time that atoms have
had to recombine, but also due to the evolution of cosmological structure,
which leads to a clumpier gas distribution, thus increasing the recombination
rate. This increase occurs even though $\bar{n}_{\rm rec}$ is continuously
``diluted'' by averaging the total number of recombinations over ever larger 
ionized fraction. 
\begin{figure*}
\begin{center}
  \includegraphics[width=3in]{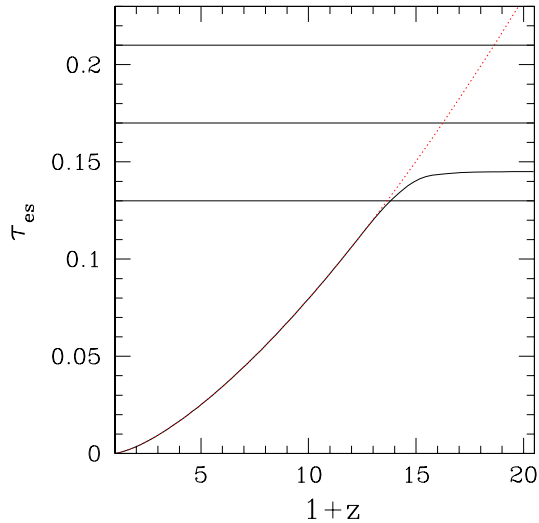}
\caption{Integrated optical depth due to electron scattering $\tau_{\rm es}$
  vs. redshift $z$ (solid). Top (dotted) curve shows the optical depth produced assuming
  complete ionization out to the corresponding redshift. Horizontal lines
  indicate the best-fit and $1-\sigma$ uncertainties of the first-year WMAP
  result, $\tau_{\rm es}=0.17\pm0.04$.
\label{tau_fig}}
\end{center}
\end{figure*}
The number of recombinations per atom in our simulation is somewhat low due to
the relatively coarse resolution of our radiative transfer grid, which effectively
filters the small-scale fluctuations. 
In Figure~(\ref{ceff_fig}, right) we plot the effective gas clumping factor in
the ionized regions, defined the usual way, 
$C_{\rm eff}=\langle n_{HII}^2\rangle/\langle n_{HII}\rangle^2$. 
Similarly to $\bar{n}_{\rm rec}$, to which it is related,
the effective gas clumping starts quite low, and grows quickly after $z\sim15$
as H~II regions start to encompass both high- and low-density regions and
cosmological structure formation progresses, reaching 1.25 towards the end of 
our simulation. The mean clumping factor is always close to one due to
the large  scales our simulation is probing, at which the density
fluctuations are relatively small. The effect of this on reionization
should be studied further, by running higher-resolution simulations
and/or including additional gas clumping at sub-grid scales.

\subsubsection{Thomson optical depth}

\begin{figure*}
\begin{center}
  \includegraphics[width=2.in]{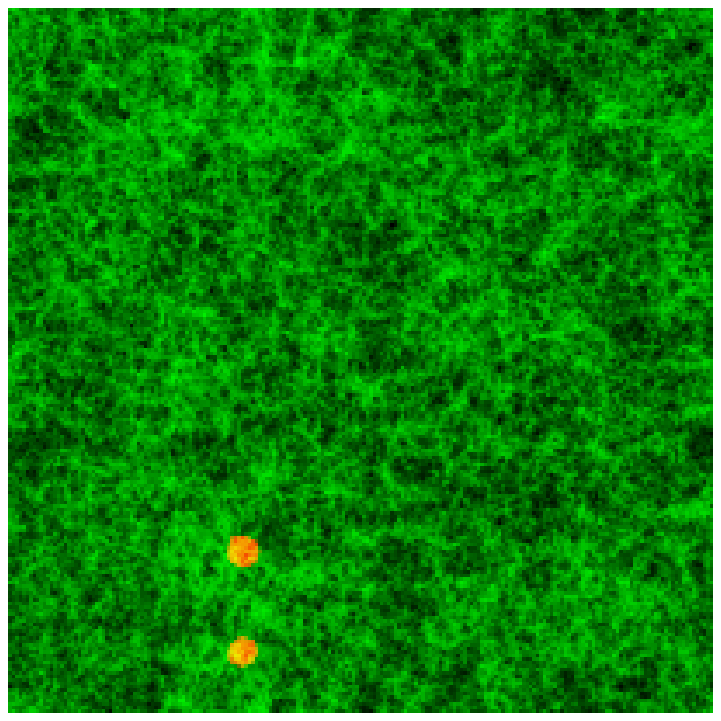}
  \includegraphics[width=2.in]{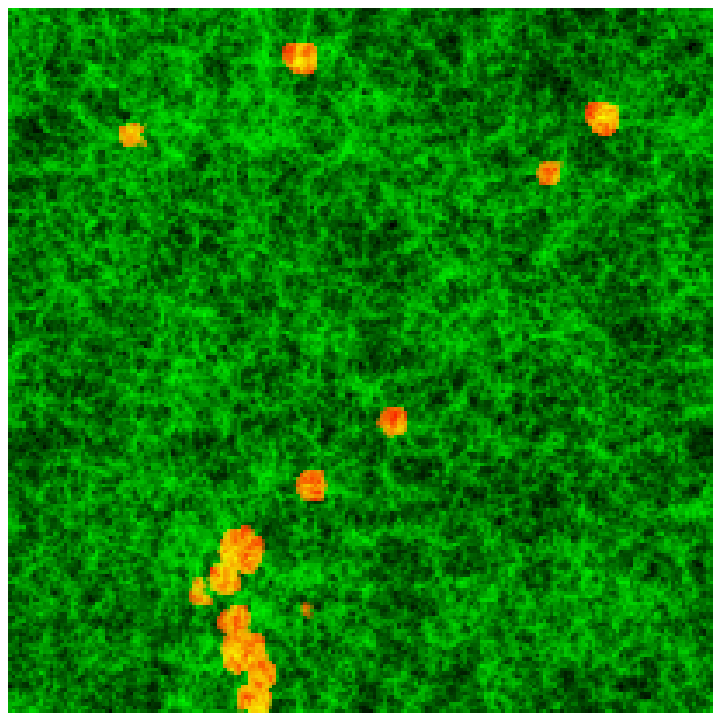}
  \includegraphics[width=2.in]{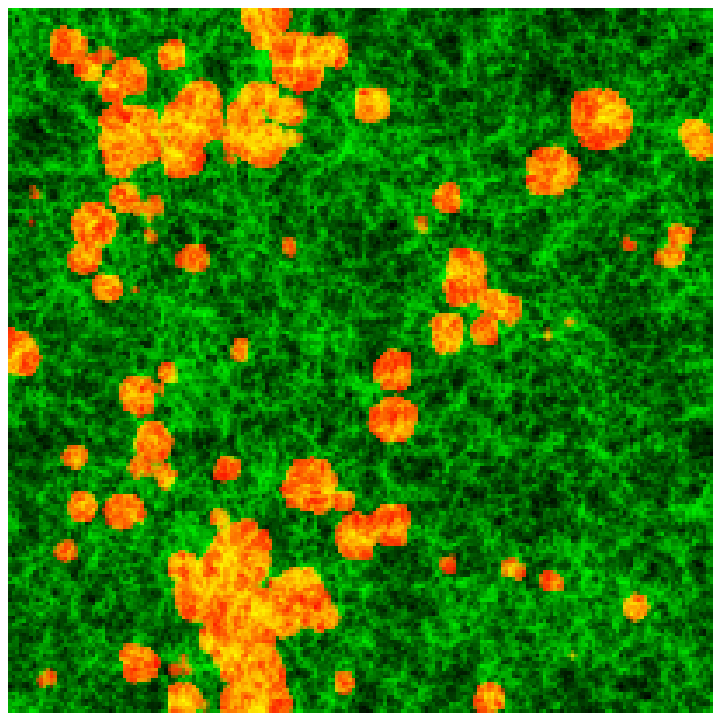}
  \includegraphics[width=2.in]{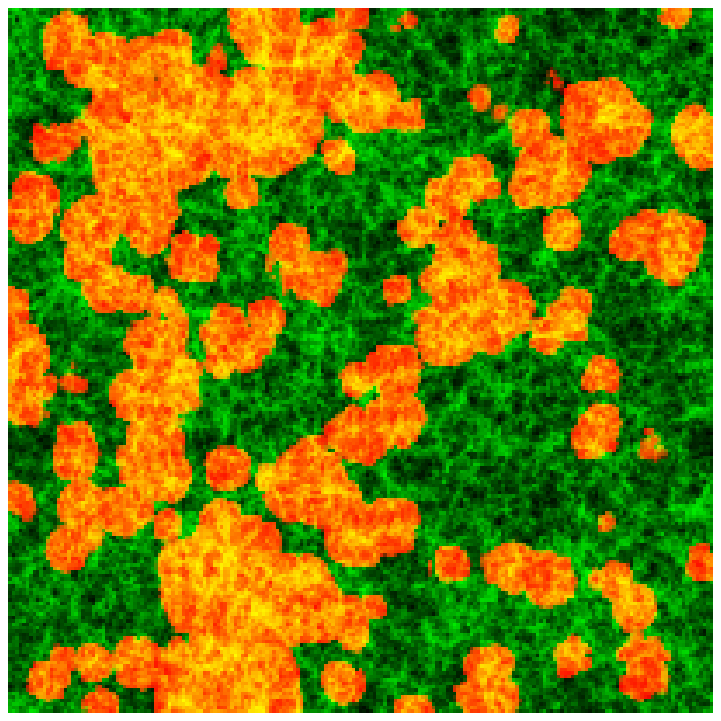}
  \includegraphics[width=2.in]{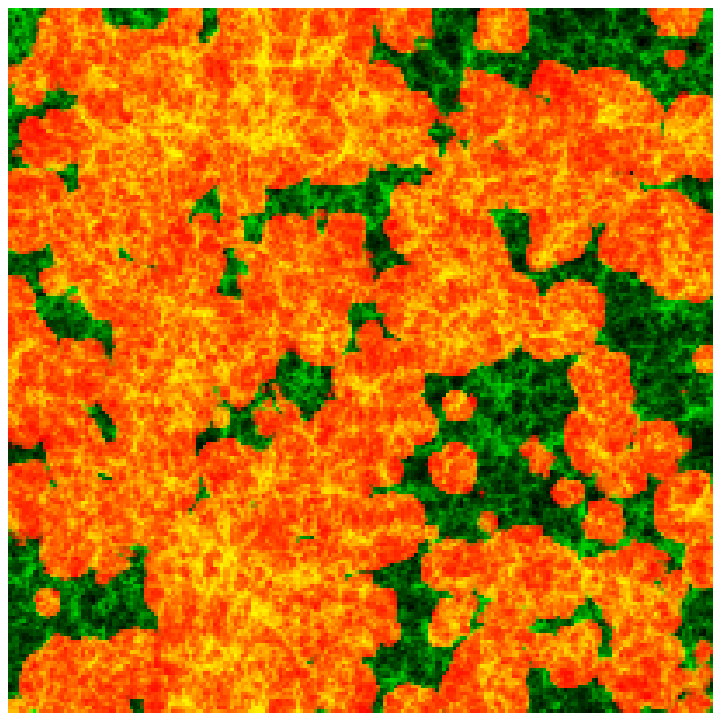}
  \includegraphics[width=2.in]{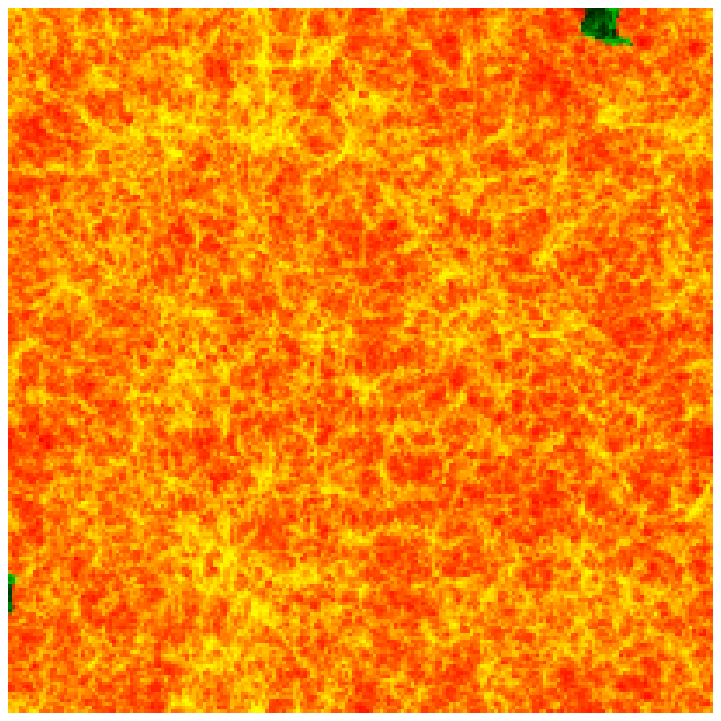}
\caption{Slices through the simulation volume at redshifts $z=18.5, 16.1, 14.5, 
  13.6$, $12.6$ and $11.3$. Shown are the density field (green in neutral regions,
  yellow in ionized regions) and the H~II regions (red). See 
$\tt www.cita.utoronto.ca/\!\sim\!iliev/research.html$ for animations of this and 
other cuts through our simulation volume. For reference the simulation box angular 
size on the sky at this redshift range is from $\sim45'$ (at $z=18.5$) to  $\sim49'$ 
(at $z=11.3$).
\label{reion_geometry_fig}}
\end{center}
\end{figure*}

For any given reionization history, the mean optical depth along a
line-of-sight between an observer at $z=0$ and a redshift $z$ due to Thomson
scattering by free electrons in the post-recombination universe is an
integrated quantity, given by
\be 
\tau_{\rm es}(z)=c\sigma_T \int^0_{z} dz' n_e(z') \frac{dt}{dz'},
\label{tau_general}
\ee 
where $\sigma_T=6.65\times10^{-25}\rm cm^{2}$ is the Thomson scattering
cross-section, $c$ is the speed of light, and $n_e(z)$ is the mean number
density of free electrons at redshift $z$, given by \be
n_e(z)=n_{H}^{0}(1+z)^3x_m(z),
\label{n_ez}
\ee where $n_H^0$ is the mean number density of hydrogen at present. We ignore
the presence of helium here, which has only a small effect on the
total electron scattering optical depth. For comparison with the value
of $\tau_{\rm es}$ between us and the surface of last scattering
inferred from measurements of the polarization of the CMB, we evaluate
$\tau_{\rm es}(z)$ at $z=z_{\rm rec}$, the redshift of recombination,
integrating over our simulation data. 

If we assume that $x_m=\rm const$ between us and a given redshift $z$ (e.g.
for an IGM fully ionized since overlap $x_m=1$) the integral in
equation~(\ref{tau_general}) has a closed analytical form
\citep{2005ApJ...624..491I} 
\be 
\tau_{\rm
  es}(z)=\frac{2c\sigma_T\Omega_b\rho_{\rm crit,0}}{3H_0m_p\Omega_0} x_m
\left\{[\Omega_0(1+z)^3+\Omega_\Lambda]^{1/2}-1\right\}.
\label{tau_ionized}
\ee 
For $x_m=1(0)$ for $z\leq z_{\rm ov}$ ($z> z_{\rm ov}$), $\tau_{\rm
  es}\geq 0.035$ for $z_{\rm ov}\geq6$, where $z_{\rm ov}$ is the redshift of
overlap. We used this analytical expression to evaluate the integrated electron
scattering optical depth between the present and the overlap, where our
simulation stops.
In Figure~(\ref{tau_fig}) we show the integrated electron scattering
optical depth we obtained. Our derived value, $\tau_{\rm es}=0.145$
is well within the $1-\sigma$ limits from the most probable value found from the
first year WMAP data. 

\subsection{The geometry of reionization}
\label{geometry_sect}

\begin{figure*}
\begin{center}
    \includegraphics[width=2.95in]{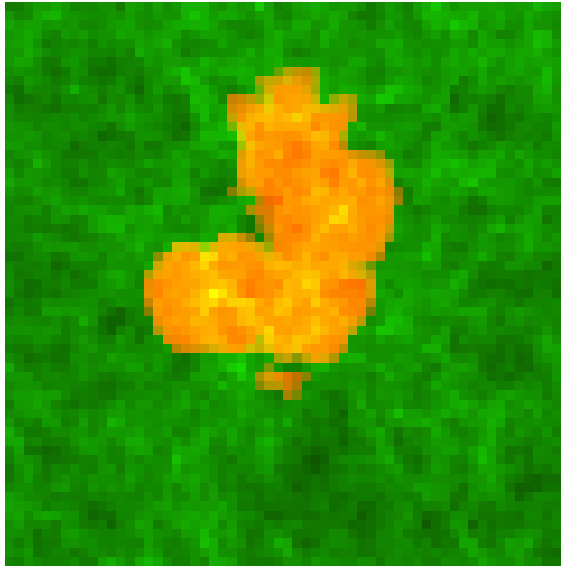}
\caption{Zoomed-in image of an early H~II regions at redshift $z=16.76$.
We used a higher-resolution ($406^3$ grid) version of the same simulation 
to show more detail. 
Note the complex, non-spherical shape. It is caused partly by the slower 
I-front propagation along the filaments and sheets of the cosmic web and
partly by merging several H~II regions.
\label{butterfly_fig}}
\end{center}
\end{figure*}

\begin{figure*}
\begin{center}
  \includegraphics[width=3in]{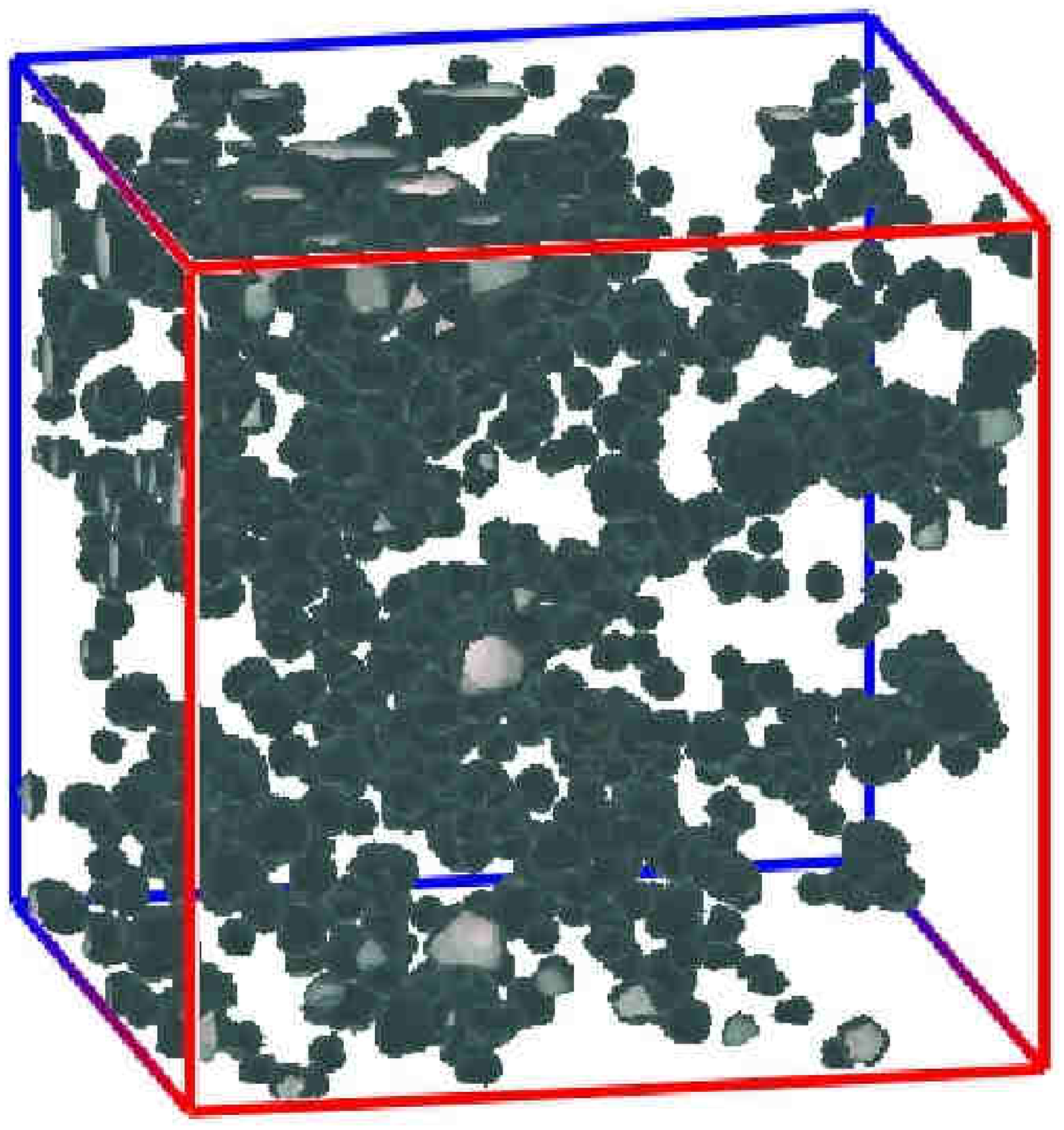}
  \includegraphics[width=3in]{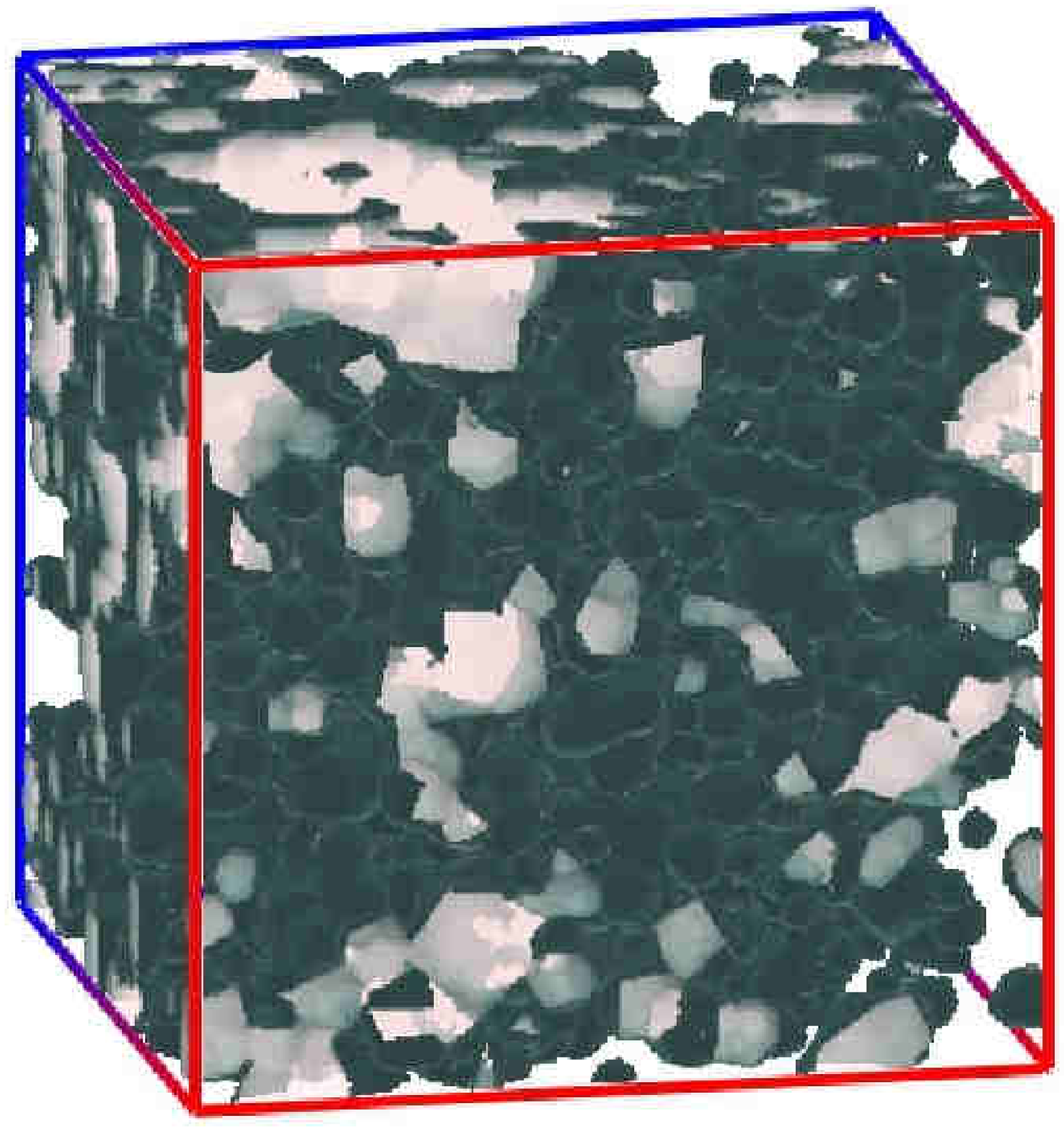}
\caption{Volume rendering of the H~II regions at redshifts $z=14.74$ (left) and 
$z=13.62$ (right). The 50\% ionization iso-surfaces are shown in dark color, while 
the light color volume-renders the ionized gas density.
(Images produced using the IFRIT visualization package of N. Gnedin)
\label{reion_geometry_3D_fig}}
\end{center}
\end{figure*}
We start our discussion of the large-scale geometry of reionization by
examining a sequence of cuts through our simulation box, shown in
Figure~\ref{reion_geometry_fig}. The first H~II regions appear in our
simulation quite early, at $z\sim21.5$ and they expand fast to a few comoving
Mpc by redshift $z\sim18$. These first H~II regions are roughly spherical,
although close examination reveals that they have more complex shapes, partly 
dictated by the slower I-front propagation along the filaments and sheets of 
the cosmic web and partly by merging several H~II regions. A zoomed-in sample
is shown in Fig.~\ref{butterfly_fig}). However, filaments cover only a very 
small fraction of each source's sky, and this fact, in addition to the 
relatively coarsely-resolved density field, results in a roughly spherical 
geometry of the isolated H~II regions. These first ionizing sources are 
highly clustered, and hence the H~II regions do not stay isolated for long 
and quickly merge together into larger and quite-irregularly shaped ionized 
regions. Hence, early-on (redshifts $z=16.1$ and $z=14.5$ in 
Figure~\ref{reion_geometry_fig}) the geometry of reionization is dominated by 
the local source clustering at the highest density peaks and along 
dense filaments. As reionization progresses ($z\sim13-14$) many more sources
form and they become less clustered.  By then both the volume and mass 
ionized fractions are about a half and there are 1-3 large regions, of sizes 
tens of Mpc, which resulted from the mergers of a number of smaller ionized 
bubbles. This creates a local overlap, while at the same time similar-size 
neutral H~I regions exist as well. Thereafter the number of ionizing sources 
in the box continues to grow strongly as they become more common, reaching tens 
of thousands in our volume. The large H~II regions continue to percolate 
locally, creating ever larger ionized zones with quite complex shapes and 
structures. At the same time there are a number of smaller H~II regions 
appearing around newly-formed sources. Eventually the whole box becomes ionized 
at $z\sim11.3$. In Figure~\ref{reion_geometry_3D_fig} we show a 3D volume 
rendering of the H~II regions at redshifts $z=14.74$ and $z=13.62$ to give the 
reader a better sense of their distribution in space and their complex shapes. 

\subsubsection{Correlation between density and ionization}

\begin{figure*}
\begin{center}
\hspace{-5mm} \includegraphics[width=2.2in]{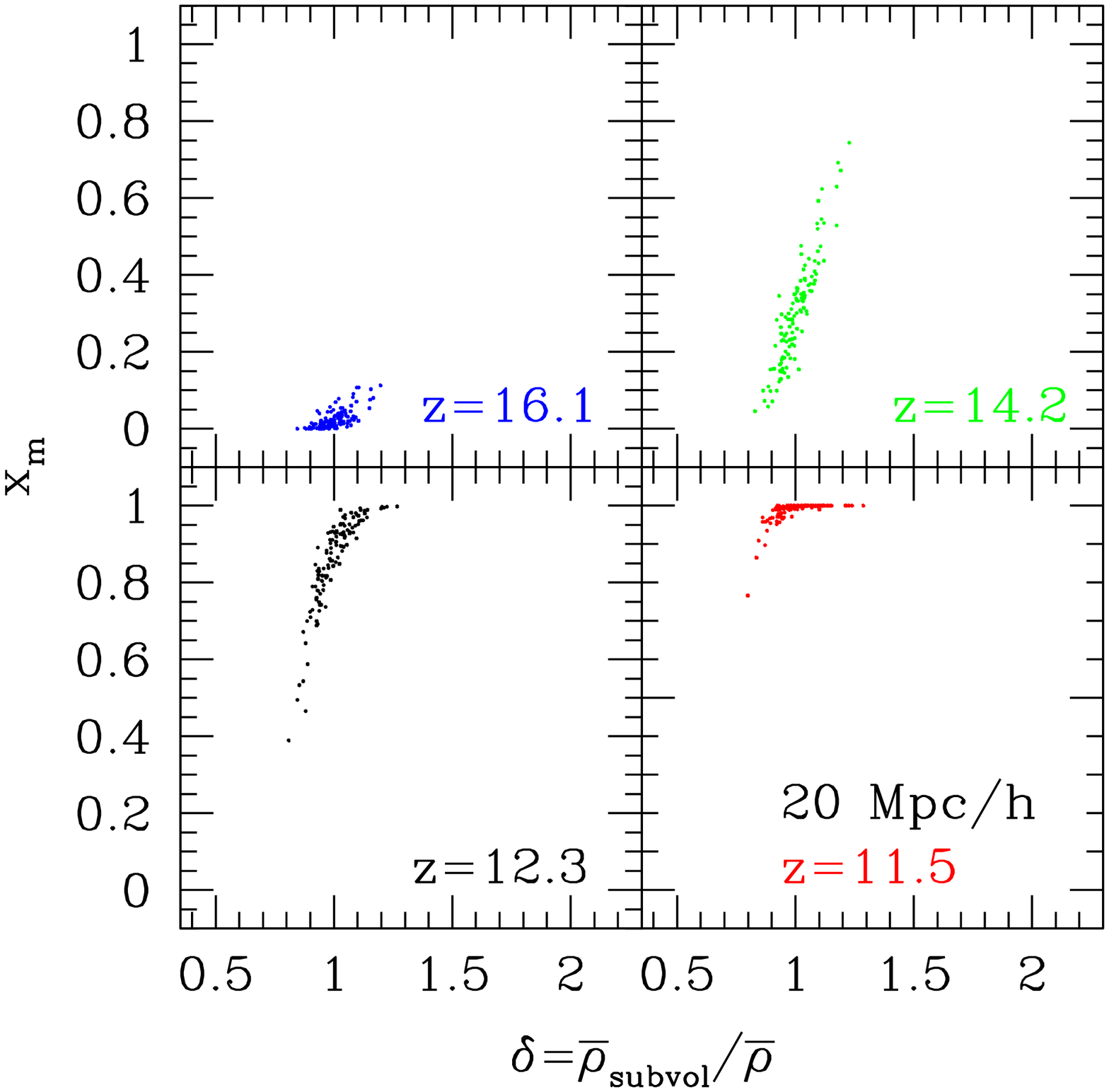}
 \hspace{-3mm} \includegraphics[width=2.2in]{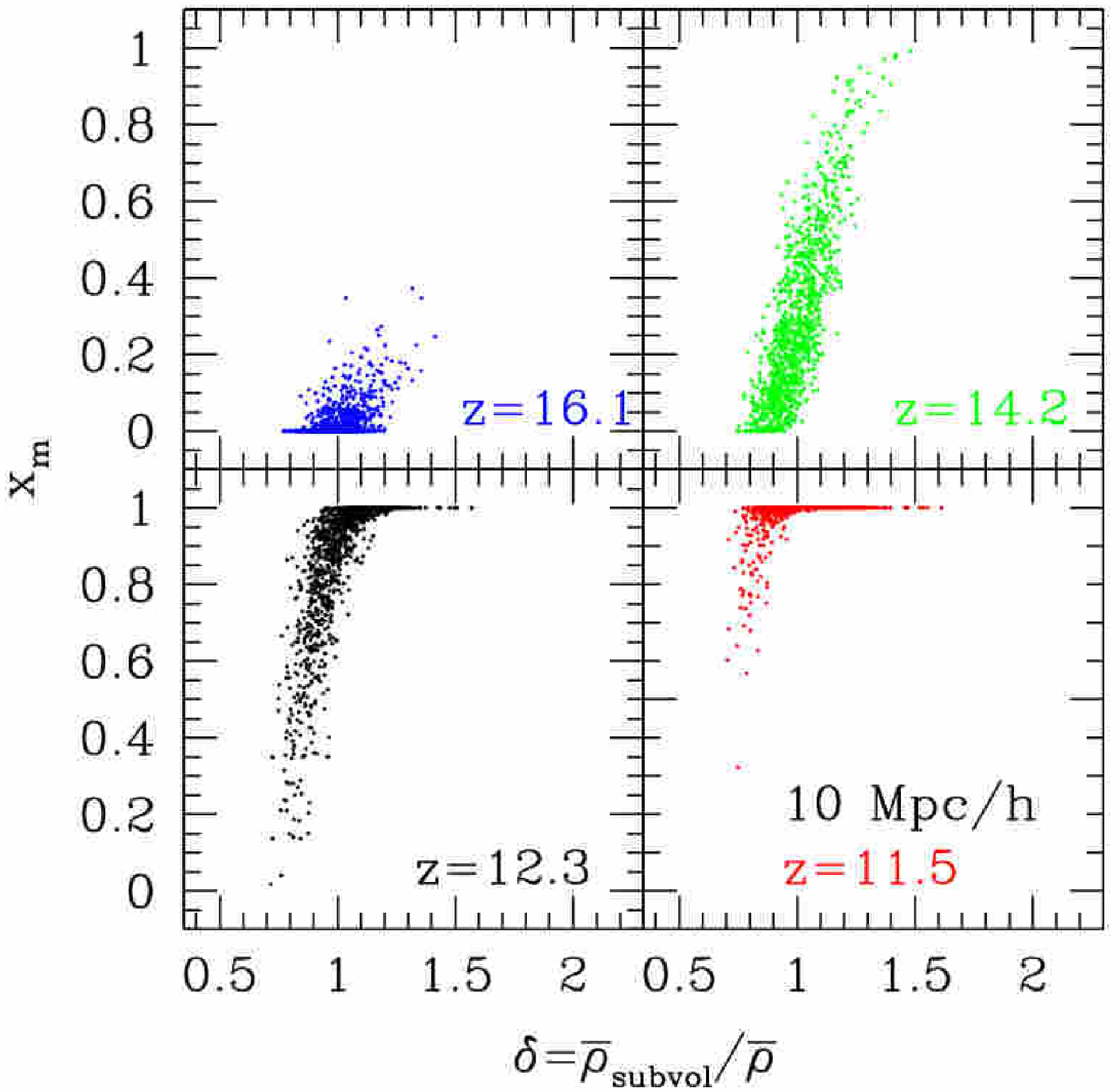}
 \hspace{-3mm} \includegraphics[width=2.2in]{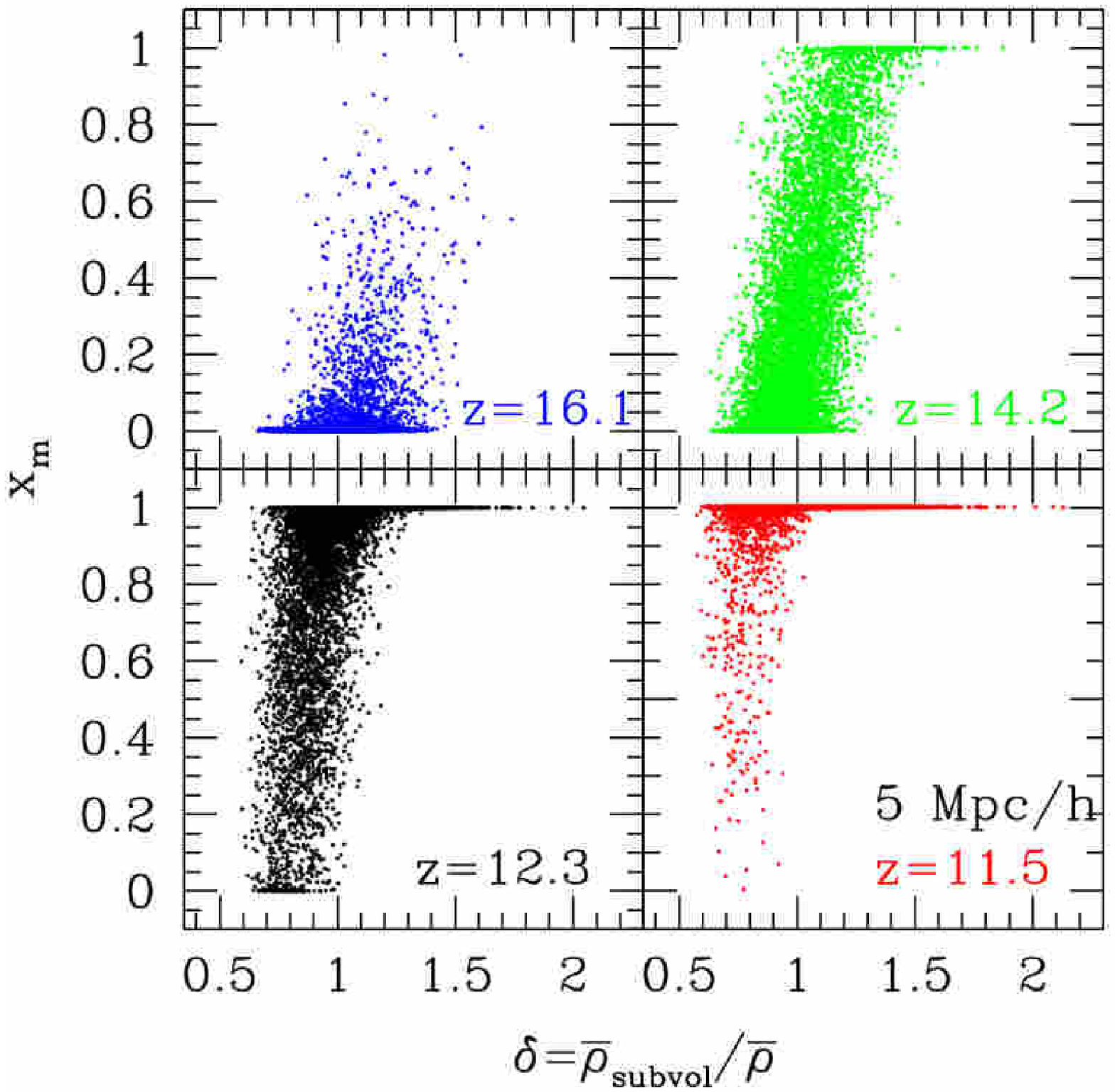}
\caption{Correlation of the mass-weighted ionized fraction $x_m$ in all 
  non-overlapping cubical sub-regions of size $20\,h^{-1}$ Mpc (left), 
  $10\,h^{-1}$ Mpc (middle) and $5\,h^{-1}$ Mpc (right) in
  our simulation volume with the average density of that region (in units of
  the mean density of the universe), 
  $\delta=\langle{\rho}\rangle/\bar{\rho}$  at several redshifts, as labeled. 
\label{del_xm_corr_fig}}
\end{center}
\end{figure*}
\begin{figure*}
\begin{center}
  \includegraphics[width=5in]{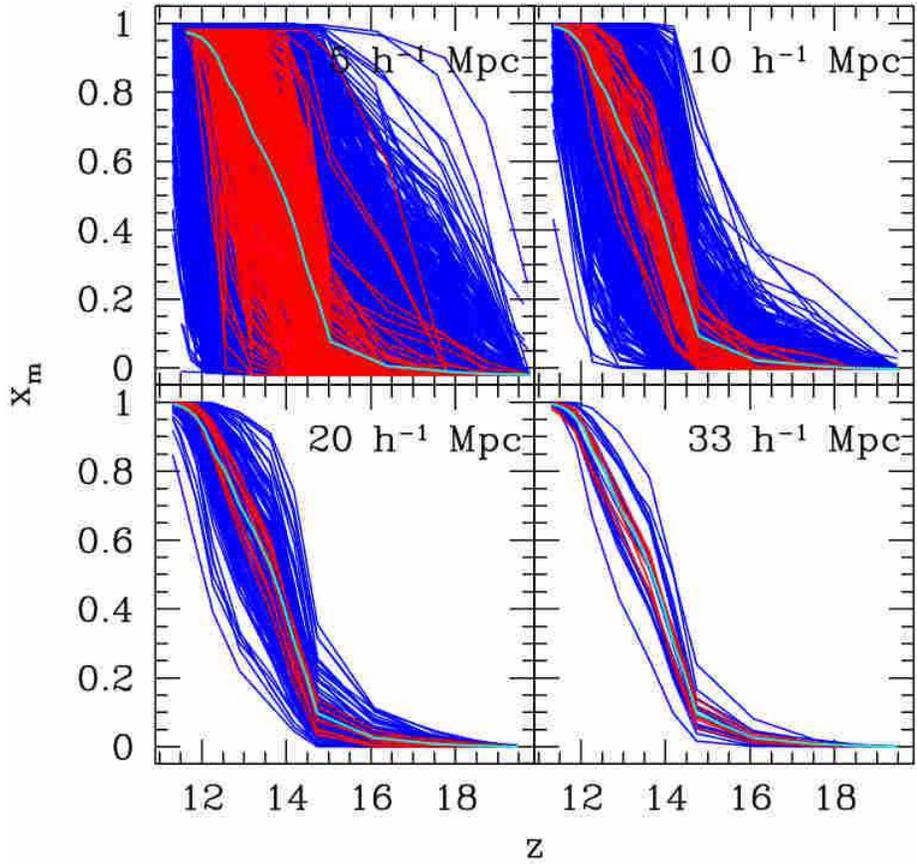}
\caption{Evolution of the mass-weighted ionized fraction $x_m$ (shown in blue)
  in all non-overlapping cubical sub-regions of sizes $5\,h^{-1}$ Mpc (top left), 
  $10\,h^{-1}$ Mpc (top right), $20\,h^{-1}$ Mpc (bottom left) and  
  $33\,h^{-1}$ Mpc (bottom right), and the sub-group of these which are at the 
  mean density (red). Also shown is the global evolution of the mass ionized 
  fraction (cyan) (the same as in Fig.~\ref{fracs_fig}).
  Even for mean-density sub-regions there is large scatter if the regions
  are smaller than few tens of Mpc. The magnitude of this scatter is 
  $\Delta z\sim1-1.5$ for the 20 Mpc regions, $\Delta z\gtrsim2$ for the 10 
  Mpc regions and $\Delta z\sim3-4.5$ for the 5 Mpc ones.
\label{mean_dens_evol_fig}}
\end{center}
\end{figure*}
\begin{figure*}
\begin{center}
  \includegraphics[width=3.2in]{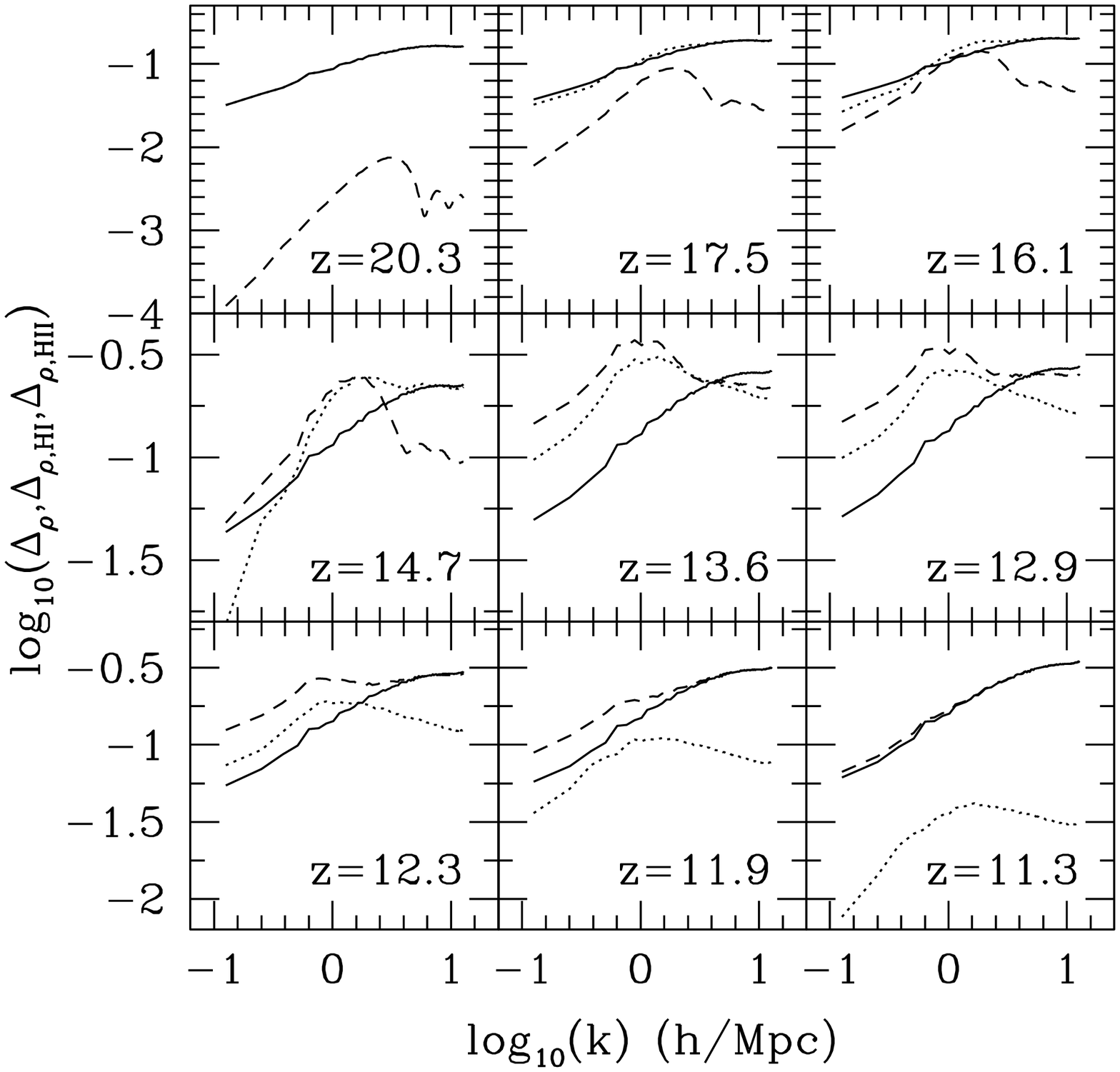}
  \includegraphics[width=3.2in]{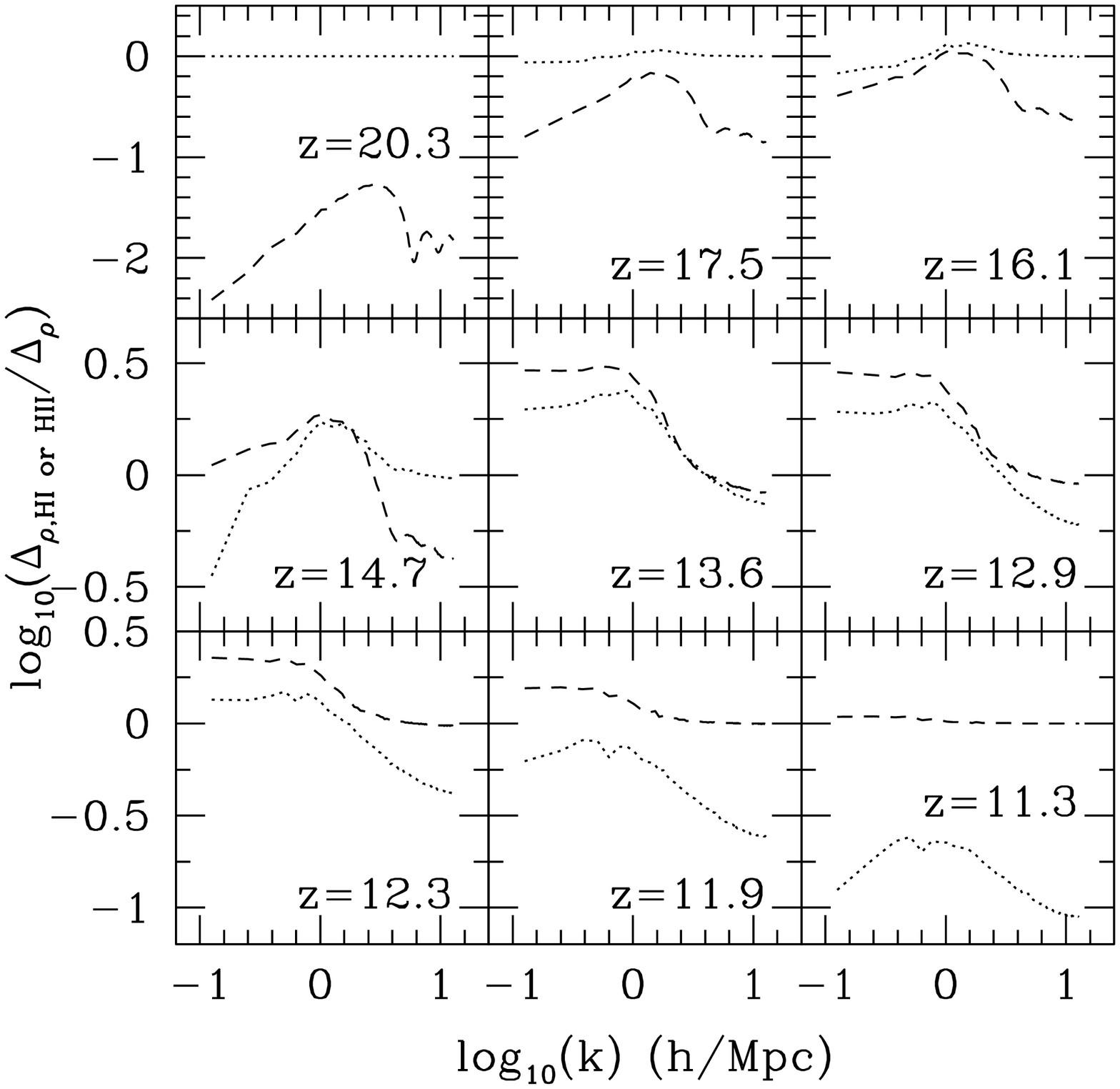}
\caption{(left) The variance, $\Delta$ of the 3D power spectra of the neutral 
  gas density (dotted), total density (solid) and ionized gas density (dashed) 
  all normalized to the total, at various 
  redshifts, as labeled. (right) Ratios of the power spectra of the neutral
  density to the total density, $\Delta_{\rho,\rm HI}/\Delta_{\rho}$, and of
  the ionized density to the total density, 
  $\Delta_{\rho,\rm HII}/\Delta_{\rho}$. 
\label{pow3d_fig}}
\end{center}
\end{figure*}
 
We found a clear but complicated correlation between the average density of a 
region in space and its reionization history. This is not unexpected since we 
have already shown above that denser regions are ionized earlier than the
less dense ones. The fact that reionization history of a region depends on the 
mean density of that region has been previously pointed out in 
\citep{2003MNRAS.343.1101C}, where the reionization histories of an overdense 
region of size $10\,h^{-1}$~Mpc and a mean-density one of size  $20\,h^{-1}$~Mpc
have been compared. 
The correlation we find has significant scatter and is strongly 
dependent on the redshift and the size of the regions considered. In 
Figure~\ref{del_xm_corr_fig} we show a scatter plot at several different
redshifts of the ionized mass fraction of all non-overlapping cubical
sub-regions 
of sizes $5\,h^{-1}$ Mpc, $10\,h^{-1}$ Mpc and $20\,h^{-1}$ Mpc vs. their
respective average density (in units of the mean density). The
overdensity-ionized fraction correlation is most clear for the larger,
$20\,h^{-1}$ Mpc regions, albeit still with significant scatter. More
importantly, the slope of the mean correlation gets steeper with time, and its
shape changes as well. The most overdense regions of that size become
completely ionized by redshift $z=13.6$ and drop from the correlation, and by
$z=11.5$ all sub-regions with mean density or above do the same.

For the smaller, $10\,h^{-1}$ Mpc sub-regions the range of densities
is much larger and the correlation still exists but with even larger
scatter. Again we notice that the mean correlation grows steeper in time, to the
point of almost disappearing at $z=11.5$ close to overlap. For the small,
$5\,h^{-1}$ Mpc sub-regions these trends become even more pronounced,
with a huge scatter and almost vertical (i.e. no correlation) mean
relation starting from $z\sim12$. 

These results clearly show that no simple relation
between the mean density of a region in space and its reionization
history exists. The two quantities are clearly
correlated, except for small regions at late times, but with
significant scatter and a mean behavior that depends on redshift and
the region's size. For smaller regions at late times the correlation
essentially disappears, with the densest  
regions completely ionized and the less dense regions at all different 
ionization stages independent of their mean density.  

This complex behavior of the sub-region reionization histories is further
illustrated in Figure~\ref{mean_dens_evol_fig}, where we show the evolution 
of the mass-weighted ionized fractions, $x_m$ for all sub-regions of the 
same size (for sizes $5\,h^{-1}$ Mpc, $10\,h^{-1}$ Mpc, $20\,h^{-1}$ Mpc and 
$33\,h^{-1}$ Mpc, as labeled), as well as of the sub-group of these which are 
within 1\% of the mean density of the universe. Also indicated is 
the evolution of the mass ionized fraction for the total simulation volume. 
The evolutionary tracks vary greatly for the small, $5$ and $10\,h^{-1}$
Mpc regions, with scatter in redshift at a given ionized fraction $x_m$ of 
up to $\Delta z\sim5$. This scatter is somewhat smaller, $\Delta z\sim1-1.5$,
but is still significant for the larger, $20\,h^{-1}$ Mpc sub-regions and
essentially disappears for the $33\,h^{-1}$ Mpc mean-density sub-regions, 
although there is still some scatter for over- and under-dense regions. The 
global ionized fraction evolution is reasonably well-represented only by the
mean-density $33\,h^{-1}$ Mpc and $20\,h^{-1}$ Mpc sub-regions.

There are a couple of reasons for this scatter and the complex behavior of the
density-ionized fraction relation. The mean density of a region in 
space is only an approximate indicator of both the number of sources inside 
that region and of the level of the density fluctuations (which dictate the
local recombination rate). Additionally, the radiative input is non-local, i.e.
sources outside a given region can, and do, influence its evolution. Both of 
these effects are much stronger for smaller regions and practically disappear 
for the $33\,h^{-1}$ Mpc regions since these are large enough to reproduce the 
mean behaviour of the universe at large. E.g. external sources matter less for
large regions since at large distances there is sufficient optical depth to
minimize their effect and limit it to the outer edges of the region. This 
clearly demonstrates that small-box reionization simulations are subject to 
a large cosmic variance, result in a range of different reionization 
histories, and cannot be used to determine precisely the redshift of overlap. 
Simulation boxes of at least a few tens of Mpc are required for any
precision and simulations with box sizes smaller than $\sim10$ Mpc
contain essentially no information about the redshift of overlap.

\subsubsection{Power spectra of the H~I and H~II regions}

The 3D power spectrum $\Delta(k)$ of a density field $\delta({\bf x})$ (in units 
of the mean) is given by 
\ba
\Delta^2(k)\equiv\frac{k^3}{2\pi^2V}\sigma_k^2=
\frac{k^3}{2\pi^2V}\int \delta^2({\bf x})e^{-i{\bf k}\cdot{\bf x}}d^3x\nonumber\\
=\frac{k^3}{2\pi^2V}\langle\delta(k)\delta(-k)\rangle,
\label{power_spec}
\ea 
where $V=L^3$ is the associated volume and ${\bf k}=(2\pi/L){\bf n}$ is
the wavenumber, where ${\bf n}$ is an integer vector. If in
equation~(\ref{power_spec}) we identify the field $\delta({\bf x})$ with 
the field of the ionized gas density fluctuations, given by $x\delta({\bf
  x})$, or the neutral gas density fluctuations, given by $(1-x)\delta({\bf
  x})$, we obtain the power spectra corresponding to these fields. In
Figure~\ref{pow3d_fig} (left) we plot the results for the 3D power spectra of 
density fluctuations, $\Delta_{\rho}$ (solid), neutral gas density
fluctuations,  $\Delta_{\rho,\rm HI}$ (dotted), and ionized gas  
density fluctuations, $\Delta_{\rho,\rm HII}$ (dashed) at redshifts $z=20.3,
17.5, 16.1, 14.7, 13.6, 12.9, 12.3, 11.9,$ and 11.3, as indicated, which cover
the complete range of our simulations. Early-on most of the gas is neutral,
and thus the neutral gas density power spectrum closely follows the total gas 
density one. As reionization progresses, its intrinsic patchiness causes both 
the neutral and ionized gas density fluctuations to rise well above the ones 
of the total gas density for wavenumbers around and below the wavenumber 
corresponding to the typical patch size at that time. The wavenumbers for which 
the fluctuations are strongly increased are roughly independent of redshift, 
$k\lesssim {\rm few}\,\rm h Mpc^{-1}$ (i.e. scales larger than few comoving Mpc). 
This strong boost in power, by a (scale-dependent) factor of up to $\sim3$ (we 
show the ratios of $\Delta_{\rho,\rm HI}$ and $\Delta_{\rho,\rm HII}$ to 
$\Delta_{\rho}$ in Figure~\ref{pow3d_fig}, right), remains in effect for the 
neutral gas density field until $z\sim12$, and for the ionized gas density 
field until the overlap epoch, after which the ionized and total density 
fluctuations power spectra of course coincide. At late times the neutral gas 
density fluctuations decrease well below the total density ones due to the 
very small remaining neutral fraction.

\begin{figure*}
\begin{center}
  \includegraphics[width=6in]{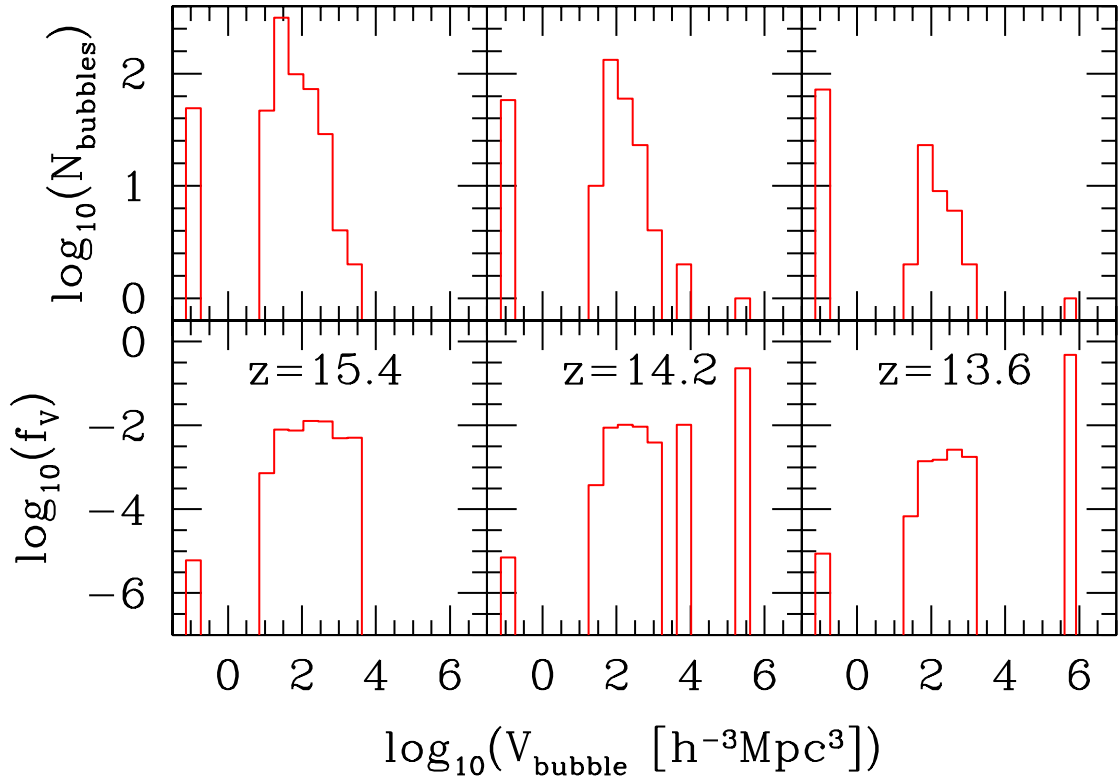}
\vspace{-1.6in}
\caption{Size distribution of ionized regions: histograms vs. bubble volume, 
$V_{\rm bubble}$ of (top) Number of H~II bubbles of that size, and (bottom) 
volume-filling factor, $f_V$, of the H~II bubbles of that size.
\label{sizes_fig}}
\end{center}
\end{figure*}

\begin{figure*}
\begin{center}
  \includegraphics[width=7in]{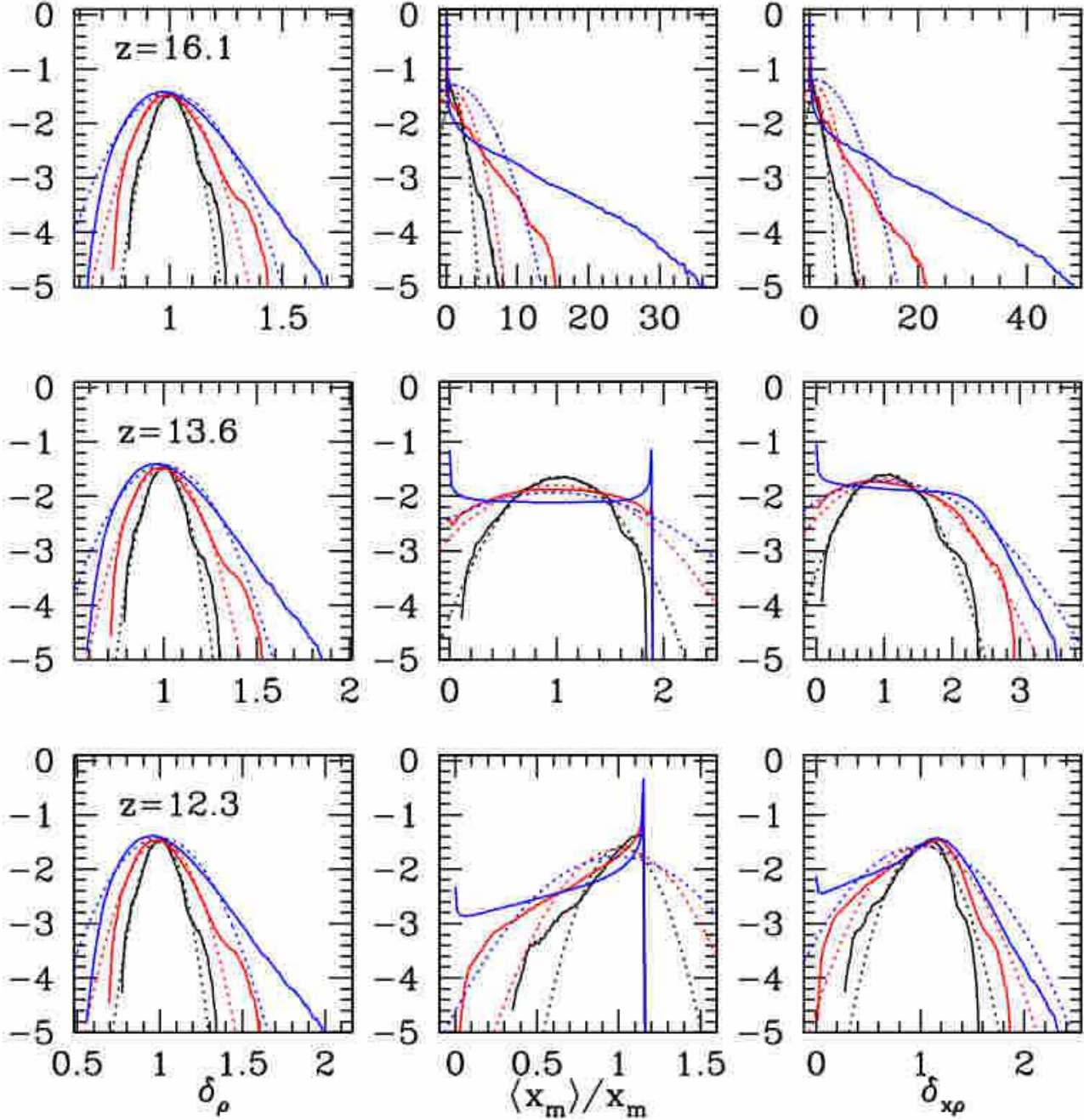}
%\vspace{-1.6in}
\caption{Non-Gaussianity of reionization: PDF distribution functions of (left 
column) the density in units of the mean density, 
$\delta_\rho=\langle\rho\rangle/\bar{\rho}$, (middle column) the mass ionized 
fraction in units of the mean one, $\langle x_m\rangle/x_m$, and (right column) 
the ionized density in units of the mean ionized density, 
$\delta_{x\rho}=\delta_\rho\langle x_m\rangle/x_m=\langle
x\rho\rangle/(x_m\bar{\rho})$, 
at redshifts $z=16.1$ (top panels), $z=13.6$ (middle panels), $z=12.3$ 
(bottom panels), for cubical regions of sizes $20\,h^{-1}Mpc$ (black solid), 
$10\,h^{-1}Mpc$ (red solid), and $5\,h^{-1}Mpc$ (blue solid). Also indicated are 
the Gaussian distributions with the 
same mean values and standard deviations (dotted, corresponding colors).
\label{gauss_fig}}
\end{center}
\end{figure*}

\begin{figure*}
\begin{center}
  \includegraphics[width=3.5in]{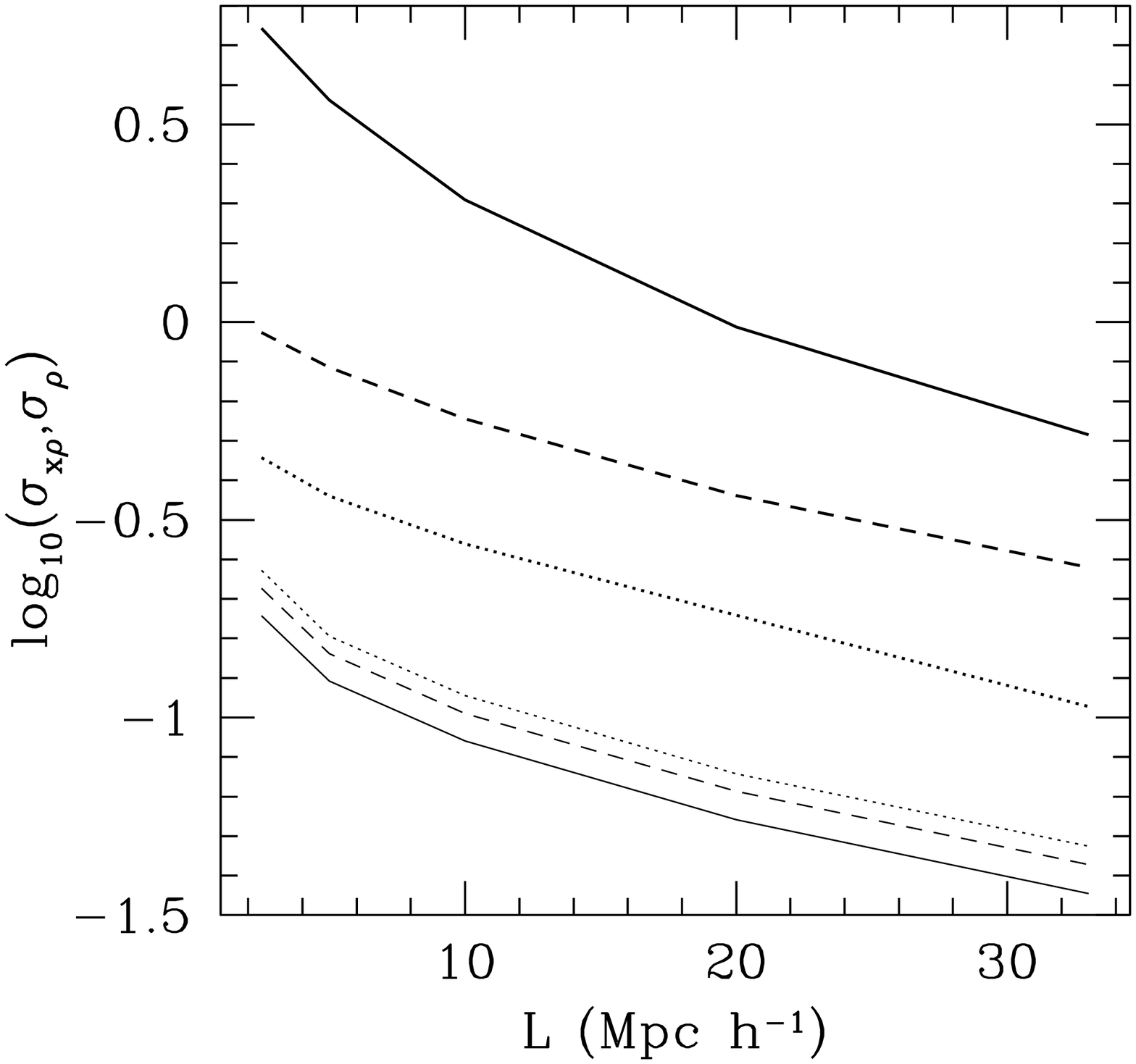}
\caption{Standard deviations (rms fluctuations) of the ionized gas density, 
$\sigma_{x\rho}$ (top, thick lines) and of the total gas density, $\sigma_{\rho}$ 
(bottom, thin lines) vs. scale at redshifts $z=16.1$ (solid), $z=13.6$ (dashed) and 
$z=12.3$ (dotted).
\label{sigmas_fig}}
\end{center}
\end{figure*}

\subsection{Size distributions of the H~II regions}

In order to find the individual ionized regions and their volumes, we employed a
friends-of-friends (FOF)-type algorithm, as follows. Two neighboring cells are
considered ``friends'' and linked together if both of their ionized fractions
are larger than 0.5. Ionized cells are thus grouped together into separate,
topologically-connected bubbles. We use the ``equivalence class'' method from
Numerical Recipes \citep{Numerical-Recipes} to determine the groups. This way 
we created catalogues of H~II bubbles and were able to calculate their numbers 
in our volume and follow their ``merger history'' in a way similar to the one 
used for finding halos in N-body simulations. It should be noted, however, that 
although there are many similarities between the two cases, as well as an obvious 
intimate connection between the source halos and the H~II regions they create,
there are also some important differences between the two distributions and merger 
histories. For example, due to the nature of the gravitational interaction,  
halos always grow in mass, while the ionized bubbles can also shrink if the total 
ionizing emissivity in the bubble decreases, due to, e.g., the death of one or 
more source. (This does not occur in the current simulations, but is expected 
to happen when small sources fall below the ionized gas Jeans mass. However, 
these are 
below our current resolution and thus not considered here). Clustering and other 
properties of these two catalogues could be expected to be quite different, as 
well. Thus, one should exercise care when making such parallels.     

Our results are shown in Figure~\ref{sizes_fig}. The total number of bubbles
starts low, at only a few, around the first sources that form in our box at
$z\sim21$. It then grows as the number of sources grows, reaching a high of
$\sim 600$ around $z\sim14-15$, still well short of the total number
of ionizing sources present then (1,500-8,000), since multiple bubbles overlap
even at these  early times. Most H~II regions have a volume of a few
tens up to $\sim10^3\,\rm(h^{-1}Mpc)^3$ 
at that time. The first large bubbles, with volumes
$>10^4\,\rm(h^{-1}Mpc)^3$, emerge 
in our simulation around $z\sim15$. Thereafter, the total number of bubbles
steadily declines, and their sizes grow as they merge, mostly with the 
largest bubbles present. By redshift $z\sim14.2$, there are three H~II regions 
larger than $10^4\,\rm(h^{-1}Mpc)^3$, and each of these occupies a few per cent or 
more of our simulation volume. By $z\sim13.6$ these percolate into one huge 
bubble which occupies about half of the total volume. There is still a large number 
of smaller H~II regions with volumes $\lesssim10^3\,\rm(h^{-1}Mpc)^3$. By redshift 
$z\sim11.5$ (close to overlap) percolation is complete and only a single 
connected bubble remains. 

In summary, our results indicate that once the ionized fraction surpasses $\sim10\%$
there are two distinct populations of H~II regions in size, a large number of 
medium-sized, $\sim10^3\,h^{-1}\rm Mpc^3$ volumes, or $\sim10\,h^{-1}$ Mpc size bubbles 
and a few, rare, very large bubbles with volumes $>10^4\rm(h^{-1}Mpc)^3$, or sizes of
tens of Mpc. (There is also a population of small, one-cell H~II regions, where the 
I-front from an ionizing source becomes temporarily trapped by its own cell.) 
In terms of ionized volume, the few large bubbles dominate the total by far, while  
the mid-size regions take only a few percent of the total. Should this behaviour
prove robust, this would have important consequences for the observability of 
reionization. The large bubbles would facilitate the direct detection of the 
ionization sources through, e.g., Lyman-$\alpha$ emitter number counts and clustering. 
Such studies would be much more difficult with the sources in the smaller H~II regions
since the source fluxes from these would be damped by nearby neutral hydrogen. Such 
very large-scale individual features would also be the easiest objects to detect with
the next generation observations in 21-cm line of hydrogen. On the other hand, 
the smaller, but numerous ionized bubbles (and neutral patches) create the large boost
of the fluctuations at intermediate scales (1'-10') of either the neutral or the 
ionized gas densities which we discussed above, demonstrated also by the power 
spectra in Fig.~\ref{pow3d_fig}. This creates a peak in the fluctuations at these 
scales, which would be the largest statistical signal from reionization, important 
for the upcoming and planned 21-cm and kSZ effect observations.

\subsection{Beyond the power spectrum: the non-Gaussian nature of reionization}

The power spectra discussed above provide an important but very limited description
of the statistics of reionization. Similar to the statistics of cosmological 
density fields, probability distribution functions (PDF) are required to provide 
more detailed description of the reionization statistics. In Figure~\ref{gauss_fig} 
we show the PDFs ($p(y)dy$, where $\int p(y)dy=1$) of the density field (left 
column), mass-weighted ionized 
fraction field (middle column), and the ionized mass field, all in units of their 
respective means, for (cubical) regions of size $20\,h^{-1}$~Mpc, $10\,h^{-1}$~Mpc, 
and $5\,h^{-1}$~Mpc at redshifts $z=16.1$ (top), $z=13.6$ (middle), and $z=12.3$ 
(bottom). We also show the corresponding Gaussian distributions with the same means 
and standard deviations. As expected, the total density PDFs are Gaussian on large, 
linear scales, and increasingly non-Gaussian at the smaller scales. At high
densities there are long non-Gaussian tails due to non-linear structure formation.

\begin{figure*}
\begin{center}
  \includegraphics[width=3.2in]{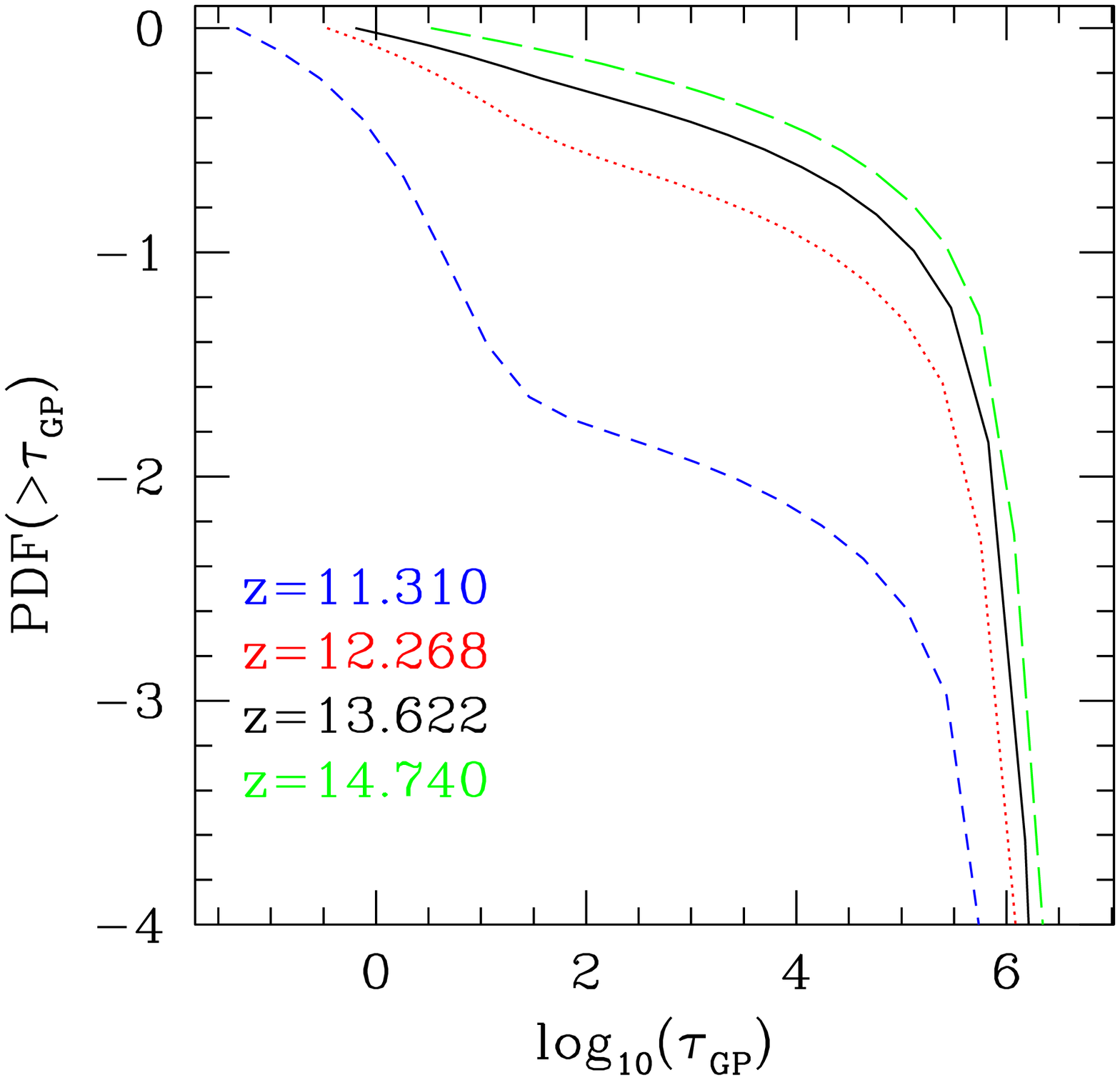}
  \includegraphics[width=3.2in]{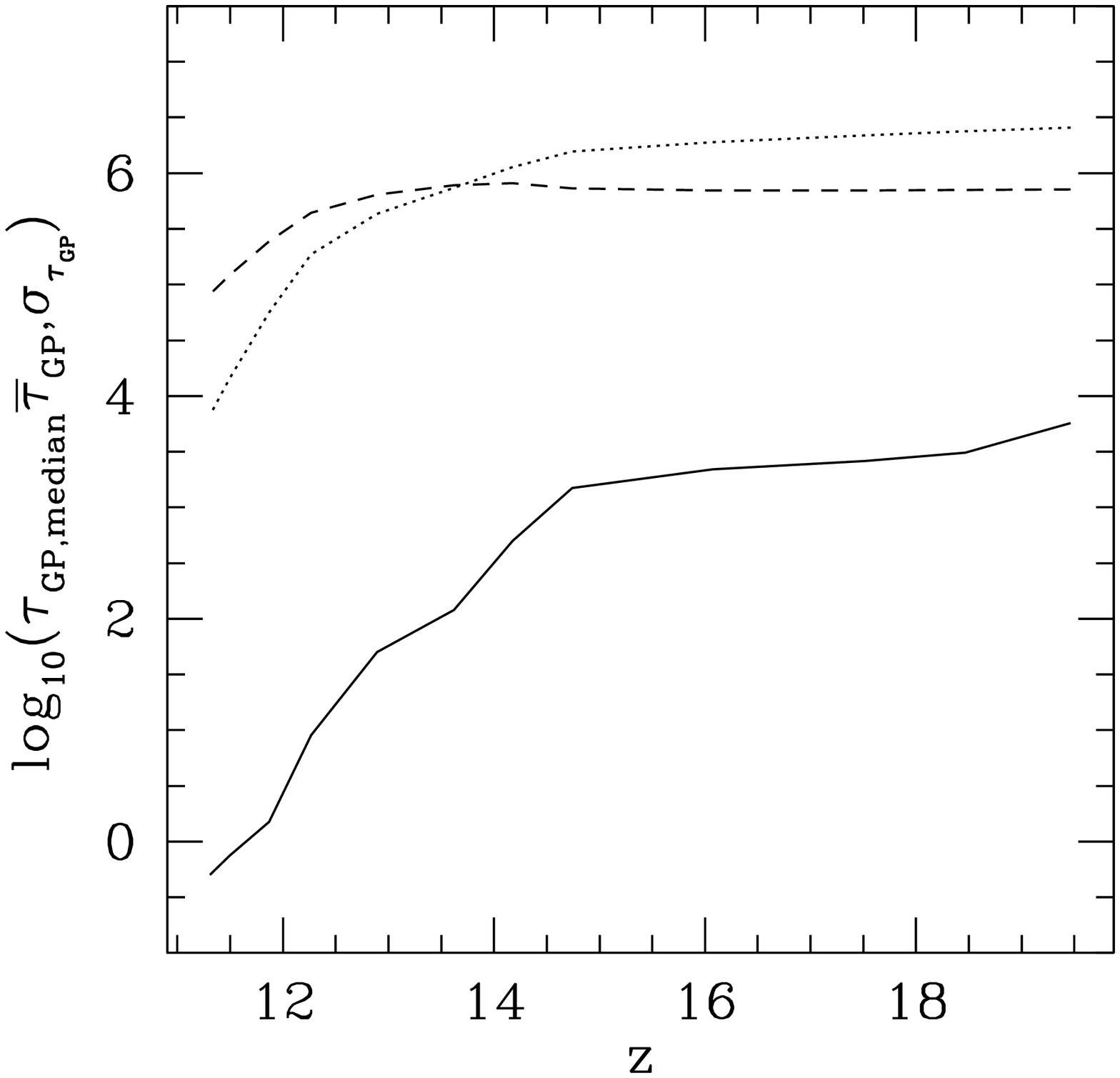}
%\vspace{-1.6in}
\caption{Gunn-Peterson optical depths: (left) Probability distribution of 
cell-by-cell Gunn-Peterson optical depths larger than a given value $\tau_{GP}$ 
at redshifts $z=14.740$ (long-dashed), $z=13.622$ (solid), $z=12.268$ (dotted) 
and $z=11.310$ (short-dashed). (right) Median value, ${\tau}_{\rm GP,median}$ 
(solid), mean value, $\bar{\tau}_{\rm GP}$ (dotted), and standard deviation, 
$\sigma_{\tau_{\rm GP}}$ (dashed), of the $\tau_{\rm GP}$ values.
\label{GP_fig}}
\end{center}
\end{figure*}  

In contrast, the mass ionized fraction distributions (middle) are generally strongly 
non-Gaussian. Early-on the distributions are strongly-peaked at 
$\langle x_m\rangle/x_m\sim0$, especially for small-size regions since the vast 
majority of them are still neutral. However, at smaller scales, which are comparable 
to the H~II region's typical sizes at that time, there is a long tail of 
highly-ionized regions well above the corresponding Gaussian. At $z=13.6$ we notice
a new feature, whereby the distribution inverts for scales below the characteristic 
size of the H~II regions, since in this case the sub-regions probed typically are 
either close to fully-ionized or to fully-neutral. At scales larger than the typical
bubble size this PDF is approximately Gaussian, but significantly wider than the 
corresponding density field PDF distribution (see also Figure~\ref{sigmas_fig})
and with a sharp cutoff at the maximum ionization fraction for a region of a given 
size, $\langle x_m\rangle_{\rm max}/x_m(z)$, where $\langle x_m\rangle_{\rm max}=1$ 
except at the largest scale (i.e. some regions of these sizes are fully-ionized).
At late times the PDFs of the ionized fraction become once again strongly non-Gaussian 
at all scales, with a long tail at small values of $\langle x_m\rangle$ and a 
sharp cutoff at $1/x_m(z)$.

Finally, we show the ionized gas density PDF in the right column of panels. Since the 
ionized density field is a convolution of the (approximately) Gaussian density field 
and strongly non-Gaussian mass-weighted ionized fraction field, the resulting
PDF exhibits features
from both and generally remains significantly non-Gaussian. At high redshifts the PDFs
generally follow the ionized fraction PDFs, but with some additional boost of the
high-values tail due to the strong correlation between high ionization and high density
(inside-out reionization) at these redshifts. At intermediate redshifts the PDF shapes
are roughly Gaussian at large scales, but wider than the corresponding density PDFs
(see also Figure~\ref{sigmas_fig}), 
and quite non-Gaussian at small scales. At late times the PDF shapes remain 
significantly non-Gaussian at all scales. The maxima of the distributions are offset 
towards the larger values compared to the mean distribution values and the decreases 
of the probabilities at smaller values of $\delta_{x\rho}$ are less steep than a 
Gaussian due to the correlation between the low-density voids and low
ionization levels discussed above.

In Figure~\ref{sigmas_fig} we show the standard deviations (i.e. rms fluctuations) 
of the ionized density field, $\sigma_{x\rho}$, and the total density field, 
$\sigma_{\rho}$, vs. scale $L$. As noted above, the ionized density distributions 
are considerably wider than the corresponding total density distributions. They 
also vary significantly more with redshift, and decrease as time goes by, unlike 
the total density fluctuations which increase with time as cosmological structures 
develop.

\subsection{Gunn-Peterson optical depths}
The observability of high-redshift Ly-$\alpha$ sources ($\lambda_\alpha=1216$\AA)
depends upon the gas ionization levels along the line-of-sight towards the observer. 
The Gunn-Peterson (GP) optical depth of the IGM at redshift $z$ is given by
\be
\tau_{\rm GP}=\sigma(\nu)n_{\rm HI}(z)(1+z)\left|\frac{dt}{dz}\right|,
\ee
where $n_{\rm HI}(z)$ is the neutral hydrogen density, 
$\sigma(\nu)=\frac{\pi e^2}{m_e}f\delta(\nu-\nu_\alpha)$ is the absorption
cross-section ($e$ and $m_e$ are the charge and mass of the electron, and $f=0.416$
is the oscillator strength of the 2p to 1s energy level transition), and 
for flat $\Lambda$CDM at high redshift, which is the relevant regime here,
\be
\left|\frac{dt}{dz}\right|=\frac{1}{H_0\Omega_0^{1/2}(1+z)^{5/2}}.
\label{dt_lam}
\ee 
Numerically, the GP optical depth at a point is then given by
\be
\tau_{\rm GP}=1.146\times10^{11}\,n_{\rm HI}(z)(\rm cm^{-3})(1+z)^{-3/2}
\ee
The probability distribution of this optical depth provides a rough idea about 
the observability of the sources in our simulation. In Figure~\ref{GP_fig} (left)
we show the probability distribution for a cell to have $\tau_{\rm GP}$ larger 
than a given value. We see that only quite close to overlap (after $z\sim12$ 
for this particular simulation) the majority of the cells become optically-thin.
In Figure~\ref{GP_fig} (right) we plot the median value, ${\tau}_{\rm GP,median}$, 
mean value, $\bar{\tau}_{\rm GP}$, and standard deviation, $\sigma_{\rm GP}$
for all cells vs. redshift. Close to overlap all three drop significantly from 
their peaks, but while the mean value and scatter remain high, the median
optical depth drops (marginally) below one. This behaviour reflects the fact
that while the majority of cells are highly ionized close to overlap, there 
exists a minority of cells which are still highly optically-thick, which
results in a high mean value and scatter for the PDF of the optical depths.
Apparently, the epoch of overlap has a significant, detectable Gunn-Peterson
optical depth throughout a substantial fraction of the volume. This is expected
since the neutral fraction in the ionized zones just prior to their overlap is 
high enough, even though it is quite small, to make $\tau_{\rm GP}\gtrsim1$ and 
it is necessary to pass the epoch of overlap before the UV background rises 
significantly due to the appearance of more distant sources \citep{1987hrpg.work..501S}.
More precise predictions for the observability of these emitters would require 
taking into account the positions of the sources and the detailed geometry of 
the ionized bubbles around them, which goes beyond the scope of the current 
paper and will be explored in a follow-up work.

\section{Summary and Conclusions}
\label{discuss_sect}
We have presented the first large-scale radiative transfer simulation of 
the reionization of the universe. The total electron-scattering optical depth 
produced by our simulation agrees well with the results on CMB polarization from
the first-year WMAP data, but
the final overlap occurs at $z\sim11$, somewhat too early to clearly agree with
the current observations of high-z QSOs and galaxies, which point to
reionization ending around $z=6-7$. However, we note that there are several 
effects which we do not include here which are expected to extend reionization 
without destroying the
agreement with the WMAP results on the optical depth. These effects include
small-scale (here sub-grid) gas clumping and lower ionizing efficiency of the
sources, among others \citep{2005ApJ...624..491I}. We are currently working 
on studying these effects in more detail with further simulations.  

For a fixed distribution of sources the I-fronts escape from the
high-density gas surrounding the sources and propagate faster into the
lower-density gas in voids. This led to predictions in earlier work that  
reionization proceeded ``outside-in'', instead, with the preferential
ionization of the lower-density regions \citep[e.g.][]{2000ApJ...535..530G}. 
We have seen some such behaviour in the toy test runs we performed as a 
simple application in \citet{methodpaper}. However, the full reionization 
simulation we have presented here points to the opposite conclusion. We 
demonstrated that in our simulation the process is ``inside-out'', i.e. 
with the high-density regions ionized earlier 
on average, and the large voids reionized last. The main reason for this is
that the character of reionization is dictated by the interplay between H~II
region expansion and evolving structure formation. The ionizing sources at
high redshifts formed at the highest-density, rare peaks of the density field.
As the cosmological structure formation progresses, more and more new sources
form inside the density peaks, as these collapse. Thus, even though the I-fronts 
due to the earliest source clusters might escape into nearby voids and ionize them
quickly, the numerous newly-formed sources ensure that on average there is always 
more mass ionized than volume. Earlier simulations which predicted
``outside-in'' reionization employed much smaller boxes, which resulted in
faster reionization and less evolution in the source population in that time 
period.

This inside-out nature of reionization leads to an increased ionizing photon
consumption since denser and clumpier gas has shorter recombination times,
resulting in multiple recombinations per atom. In our particular simulation
approximately 1.6 ionizing photons per atom were eventually required for 
completing the process.  Our conclusions are not affected by the relatively coarse 
resolution of the simulation presented here, with cell size $\sim0.5\,h^{-1}$~Mpc, 
which significantly filters the density fluctuations. We have now 
also run a simulation with the same underlying density field and sources, but with 
higher radiative transfer grid resolution ($406^3$ cells) \citep{21cmreionpaper}, 
as well as multiple simulations with smaller box size ($35\,h^{-1}$~Mpc), and thus 
higher spatial resolution \citep{selfregulated}. Increased spatial resolution and 
the corresponding higher density contrasts emphasizes the inside-out nature 
of reionization even more and only strengthtens our current conclusions.  
In turn, such increased ionizing photon consumption would require fairly 
efficient production of ionizing photons at high-z, either due to massive, hot 
stars, high efficiency of star formation, high escape fractions, or a combination 
of all these. 

Our current simulation does not include the effects of mechanical feedback (e.g. 
from supernovae) which can disrupt the smaller sources and modify their star 
formation rates. Currenly we also do not include the radiative input from ionizing 
sources with halo masses smaller than $2.5\times10^9\,M_\odot$, which are below 
the resolution of our N-body simulation. Hence, our source population should be 
considered a low limit to the realistic one. These effects would be studied in a 
subsequent work \citep{selfregulated}.

The scale of our simulation has allowed us for the first time to study numerically
the large-scale geometry and statistics of reionization. We derived the correlation
between the density of a region and its ionization state. We showed that the relation
is complex and its mean and dispersion are significantly redshift- and 
scale-dependent. At late times and small scales the correlation essentially 
disappears. Furthermore, we demonstrated that the reionization history of sub-regions
exhibits significant scatter at small scales and provides good description of the
mean behaviour only at large scales, above 20-30 $h^{-1}$ Mpc. 

The patchiness of reionization results in a significant increase of both the neutral- 
and ionized-gas density fluctuations, with important 
implications for statistical observations of reionization at, e.g., the 21-cm line of 
neutral hydrogen \citep{21cmreionpaper} and small-scale CMB anisotropies. The power 
spectra of ionized and
neutral density fluctuations at wavenumbers $k\gtrsim1\,\rm h\,Mpc^{-1}$ are boosted
by up to factors of $\sim2-3$ compared to the fluctuations of the total gas density. 
This roughly translates to a stronger fluctuation signal at arcminute and larger 
scales.

We derived the size distributions of the H~II regions in our simulations in both number
counts and volume filling factors. We found two populations of bubbles, one with a 
large numbers of medium-sized ($\sim10$ Mpc) and one with a few, rare and very large 
bubbles of size tens of Mpc. The first population provides most of the statistical
fluctuations at arcminute scales discussed above, while the large ones should be the
most prominent and most easily-detectable features of reionization. We also derived 
the distribution of Gunn-Peterson optical depths in our simulation volume as a first 
step to more detailed predictions for the observations of Ly-alpha emitters at high 
redshift by current and upcoming large surveys.

Finally, we demonstrated and for the first time quantified the non-Gaussian nature of 
the reionization statistics. The probability distribution functions for both the 
ionized mass fraction and the ionized mass are generally strongly non-Gaussian on all
scales and at all times, especially at the beginning and end of reionization. At 
scales below the typical scales for the H~II regions the distribution even inverts, 
with the highest probability for a region to be either highly-neutral or highly-ionized.
This is a feature of our reionization model, where no sources vary strongly in 
luminosity over time (or die). This is justified for our ionizing sources, which are 
relatively large galaxies above the ionized-gas Jeans mass, whose formation 
cannot, therefore, be easily suppressed. We will study the effects of the
presence of smaller ionizing sources and more realistic source behaviour in a 
follow-up work.

%\acknowledgements
\section*{Acknowledgments} 
This work was partially supported by NASA Astrophysical Theory Program grants
NAG5-10825 and NNG04GI77G to PRS. GM acknowledges support from the Royal 
Netherlands Academy of Art and Sciences. MAA is grateful for the support of 
a DOE Computational Science Graduate Fellowship.
%This work greatly benefited by the  

%\bibliographystyle{apj} \bibliography{../refs}
\bibliographystyle{mn} \bibliography{../refs}

\end{document}